\documentclass{JHEP3} 

\title{New Superembeddings for Type II Superstrings}
\author{D. V. Uvarov\\
NSC Kharkov Institute of Physics and Technology\\
61108 Kharkov, Ukraine\\
E-mail: \email{d\_uvarov@hotmail.com, uvarov@kipt.kharkov.ua}}

\abstract{Possible ways of generalization of the superembedding approach for the supersurfaces with the number of Grassmann directions being less than the half of that for the target superspace are considered on example of Type II superstrings. Focus is on $n=(1,1)$ superworldsheet embedded into $D=10$ Type II superspace that is of the interest for establishing a relation with the NSR string.}
\preprint{}
\begin{document}
\setcounter{equation}{0}
\def\theequation{\thesection.\arabic{equation}}

\section{Introduction}
The interpretation  of superbranes as solitons of supergravity/super-Yang-Mills theories for the first time suggested by J.~Hughes et.al. \cite{HughesLP} (for reviews see \cite{Soliton}) allows to connect partial breaking  of target space supersymmetries caused by these solitonic configurations  with $\kappa-$symmetry inhereit to the brane worldvolume actions. In particular, for conventional superbranes that leave unbroken half of the target space supersymmetries $\kappa-$symmetry has the same number of independent parameters. In the proper gauge the action of $\kappa-$symmetry transformations for one half of the target space Grassmann coordinates $\eta$ coincides with that of supersymmetry transformations. It is these coordinates that are associated with the unbroken part of target space supersymmetries and they can be identified with Grassmann coordinates on the superworldvolume. The other components of the target space Grassmann coordinates $\theta^{\prime}$ constitute Goldstone fermions associated with the broken supersymmetries. 

The worldvolume nature of $\kappa-$symmetry was revealed later by D.V. Volkov and collaborators \cite{VZ}-\cite{STVZ}. (It should be noted, that the early attempts to combine the worldsheet and target-space supersymmetries for particles and strings were made in the middle of 80's \cite{Gates}. However, suggested method was too general to describe conventional models. For recent developments see \cite{GL}.) This inspired the investigation of worldvolume superfield formulations of superbranes \cite{DPSF}-\cite{Drum}, where $\kappa-$symmetry is replaced by the more fundamental worldvolume superdiffeomorphism symmetry. These worldvolume superfield actions upon integration over the worldvolume Grassmann variables and the elimination of auxiliary fields by means of the equations of motion have to reduce to the Green-Schwarz-type actions, where $\kappa-$symmetry appears as a remnant of the local worldvolume superdiffeomorphism symmetry. Further gauge fixing of the local symmetries for these Green-Schwarz-type actions yields the theory that is invariant under worldvolume global supersymmetry. So the role of $\kappa-$symmetry is to assemble local symmetries gauge fixed bosonic and fermionic degrees of freedom of the brane into the worldvolume supermultiplets. This is well known bose-fermi matching. Taken as the key principle for constructing superbranes it allows to define possible values for $p$ and $D$ (dimension of the target superspace), or, the brane scan. However, it turns out that for some types of branes like Type II superstrings in $D=4,6,10$ \cite{Galperin}, \cite{BPSTV}, $D=11$ supermembrane \cite{BPSTV}, $M5-$brane \cite{BLNPST} the construction of the worldvolume superfield actions with $\kappa-$symmetry being entirely replaced by the worldvolume superdiffeomorphisms encounters significant difficulties  just in the point of the relation to the corresponding Green-Schwarz-type actions since the present in the doubly supersymmetric formulations auxiliary fields become dynamical. 

The systematical study of doubly supersymmetric brane formulations is achieved in the framework of the superembedding approach (for reviews see \cite{DS}, \cite{HSW}). This is the supersymmetric generalization of the theory of surface embeddings \cite{Eis} and deals with a brane superworldvolume embedded into a target superspace. Initially the idea of the consideration of bosonic strings as surfaces embedded into the target space was put forward in \cite{RLO} (see also \cite{BN}). The relation with the gauge theories on the worldvolume was studied in \cite{Zh81}. The first indications of the efficiency of embedding approach in the supersymmetric theories of particles and strings were discovered in \cite{VZ}-\cite{STVZ}. Further developments of superembedding approach and geometrodynamical condition as its basic ingredient were made in \cite{DGHS}. Detailed analysis of the case when the number of the superworldvolume Grassmann directions $n_{wv}$ equals half the number of the target superspace Grassmann coordinates $n_{ts}$ was performed in \cite{BPSTV}. 

In supersymmetric theory of embeddings there appears the interesting phenomenon that has no analogs in the classical bosonic theory, namely, for certain branes geometrodynamical equation contains superbrane equations of motion. This leads to the new class of on-shell embeddings that are minimal from the beginning. It is precisely for the branes of this class one faces the described above obstacles when trying to construct worldvolume superfield models. (It is possible, however, to construct worldvolume superfield actions for the gauge fixed physical degrees of freedom in the framework of the method of nonlinear realizations both for off-shell and on-shell superbranes. The examples considered in the literature include codimension zero and one cases (see \cite{Ivanov} and references therein). Superembedding approach treatment for these superbranes was performed e.g. in \cite{BPPST}, \cite{Drum}, \cite{HKSS}.)  On the other hand, for on-shell embeddings the basic superembedding equation encompasses full information about the superbrane dynamics. If the superembedding is off-shell then the superbrane equations of motion follow from the requirement of the minimality of embedding. The situation is similar to that in the bosonic theory \cite{Eis}-\cite{Zh81}. Then one is able to construct the doubly supersymmetric superfield actions for corresponding superbranes.  This scheme was realized for various superparticle models \cite{STV}, \cite{STVZ}, \cite{DPSF}-\cite{BMS}, null-superstrings \cite{BSTV}, heterotic superstrings \cite{Berkovits1}-\cite{BB}, $D=3$ Type II superstring \cite{Galperin} (see also \cite{CP}), $D=4$ $N=1$ supermembrane \cite{HRS}, \cite{PST} and the space-filling branes in $D=3,4$ \cite{BPPST}, \cite{Drum}. However, proposed worldvolume superfield actions possess infinitely reducible superfield gauge symmetries that cause significant difficulties when trying to quantize these models. Alternative approach for constructing superbrane worldvolume superfield actions is provided by the generalized action principle \cite{BSV}-\cite{BST}.

The characteristic feature of the superembeddings considered so far is that the number of the superworldvolume Grassmann directions $n_{wv}$ equals to the half of the number of target superspace Grassmann directions $n_{ts}$. This situation corresponds to the most complete description of superbranes, where $\kappa-$symmetry of component formulations is entirely replaced by the local worldvolume supersymmetry. However, it is possible to contemplate more general situation with $n_{wv}<n_{ts}/2$ (the case $n_{wv}>n_{ts}/2$ seems to be too restrictive although there exist (super)particle models \cite{BLS} with more than $n_{ts}/2$ $\kappa-$symmetries (see also \cite{ZL})) in an attempt to extend the class of off-shell superembeddings for which there exist worldvolume superfield formulations. Considered in the literature worldvolume superfield actions  for particles and strings with $n_{wv}<n_{ts}/2$ \cite{VZ}-\cite{STVZ}, \cite{Berkovits1}, \cite{Tonin}, \cite{Berkovits2}, \cite{Berkovits3}, \cite{DIS}, \cite{CP}, \cite{APT}, for which $\kappa-$symmetries that were not realized as worldvolume superdiffeomorphisms appear as the local symmetries on the superfield level, suggest that such embeddings can still describe breaking of the half of target space supersymmetries, however, there can also exist configurations that break larger fraction of supersymmetry. It therefore seems to be instructive to study such embeddings with $n_{wv}<n_{ts}/2$ in the framework of the superembedding approach. In the present paper we consider $D=10$ Type II superstrings as the representatives of on-shell (when $n_{wv}=n_{ts}/2$) superembeddings. We concentrate on $n=(1,1)$ superworldsheet embedded into the flat $D=10$ Type II target superspace that is of the most interest for establishing the relation with the NSR string \cite{VZ}-\cite{STVZ}, \cite{Berk}-\cite{me} invariant under the local $n=(1,1)$ worldsheet supersymmetry. 

\section{Basic equations}

The necessary ingredients of our construction are vector and spinor Lorentz harmonic variables \cite{GIKOS}-\cite{harmonics} $(u_{\underline m}^{(\underline n)},v^{(\underline\alpha)}_{\underline\mu})$ and their inverse $(u^{\underline m}_{(\underline n)},v_{(\underline\alpha)}^{\underline\mu})$. For the $D=10$ target superspace these are $10\times10$ and $16\times16$ orthonormal matrices respectively. Spinor harmonics also should satisfy certain harmonicity conditions that reduce the number of independent degrees of freedom of the spinor harmonics to the dimension of $SO(1,9)$ Lorentz group equal to 45. In the presence of the superstring $SO(1,9)_R$ group acting on the index in brackets of harmonics is broken down to $SO(1,1)\times SO(8)$, thus the Lorentz harmonics adapted for the description of superstrings acquire the form 
\begin{equation}\label{1}
u^{(\underline n)}_{\underline m}=\left(u^{+2}_{\underline m}, u^{i}_{\underline m}, u^{-2}_{\underline m}\right), v^{(\underline\alpha)}_{\underline\mu}=\left(v^{+}_{\underline\mu q}, v^{-}_{\underline\mu\dot q}\right), v_{(\underline\alpha)}^{\underline\mu}\equiv \left(v^{(\underline\alpha)}_{\underline\mu}\right)^{-1}=\left(v_{q}^{-\underline\mu}, v_{\dot q}^{+\underline\mu}\right)
\end{equation}
in the light-like notations for $SO(1,1)$ that is identified with the worldsheet structure group. 
Indices $i, q, \dot q$ corespond to $v, c, s$ representations of $SO(8)$. Lorentz harmonics (\ref{1}) parametrize the coset-space $SO(1,9)/SO(1,1)\times SO(8)$.
Lorentz connection 1-form looks like
\begin{eqnarray}
\Omega^{(\underline n)(\underline k)}=u_{\underline m}^{(\underline n)}du^{\underline m(\underline k)}=\left(\Omega^{(0)}, \Omega^{\pm2i}, \Omega^{ij}\right):\nonumber\\
\Omega^{(0)}=\frac12 u_{\underline m}^{+2}du^{\underline m-2}, \Omega^{\pm2i}=u_{\underline m}^{\pm2}du^{\underline m i}, \Omega^{ij}=u_{\underline m}^{i}du^{\underline m j}.\label{cartan}
\end{eqnarray}
In this set up $D\!=\!10$ Type II target-superspace coordinates $(X^{\underline m}, \Theta^{1\underline\mu}, \Theta^2_{\underline\mu})$ (IIA) or $(X^{\underline m}, \Theta^{1\underline\mu}, \Theta^{2\underline\mu})$ (IIB) can be splitted as follows
\begin{equation}\begin{array}{c}\label{1.4}
X^{(\underline n)}=X^{\underline m}u_{\underline m}^{(\underline n)}=\left(x^{+2}, x^i, x^{-2}\right),\\[0.3cm] \Theta^{1(\underline\alpha)}=\Theta^{1\underline\mu}v^{(\underline\alpha)}_{\underline\mu}=\left(\theta^{1+}_q, \theta^{1-}_{\dot q}\right), \Theta^2_{(\underline\beta)}=\Theta^2_{\underline\mu}v^{\underline\mu}_{(\underline\beta)}=\left(\theta^{2-}_q, \theta^{2+}_{\dot q}\right) 
\end{array}\end{equation}
for the type IIA superspace and 
\begin{equation}\begin{array}{c}\label{1.5}
X^{(\underline n)}=X^{\underline m}u_{\underline m}^{(\underline n)}=\left(x^{+2}, x^i, x^{-2}\right),\\[0.3cm] \Theta^{1(\underline\alpha)}=\Theta^{1\underline\mu}v^{(\underline\alpha)}_{\underline\mu}=\left(\theta^{1+}_q, \theta^{1-}_{\dot q}\right), \Theta^{2(\underline\alpha)}=\Theta^{2\underline\mu}v^{(\underline\alpha)}_{\underline\mu}=\left(\theta^{2+}_q, \theta^{2-}_{\dot q}\right) 
\end{array}\end{equation}
for the type IIB superspace.

Basic 1-forms of the Type IIA target superspace then acquire the form
\begin{equation}\label{ba}
\Pi^{\underline m}=dX^{\underline m}-id\Theta^{1\underline\mu}\sigma^{\underline m}_{\underline\mu\underline\nu}\Theta^{1\underline\nu}-id\Theta^{2}_{\underline\mu}\tilde\sigma^{\underline m\underline\mu\underline\nu}\Theta^{2}_{\underline\nu}=\frac12 u^{\underline m+2}\Pi^{-2}+\frac12 u^{\underline m-2}\Pi^{+2}-u^{\underline mi}\Pi^i,
\end{equation}
\begin{equation}\label{1.6}
d\Theta^{1\underline\mu}=v^{\underline\mu}_{(\underline\alpha)}\pi^{1(\underline\alpha)}=v^{-\underline\mu}_q\pi^{1+}_q+v^{+\underline\mu}_{\dot q}\pi^{1-}_{\dot q},
\end{equation}
\begin{equation}\label{fa}
d\Theta^2_{\underline\mu}=v_{\underline\mu}^{(\underline\alpha)}\pi^2_{(\underline\alpha)}=v^{+}_{\underline\mu q}\pi^{2-}_q+v^{-}_{\underline\mu\dot q}\pi^{2+}_{\dot q},
\end{equation}
where
\begin{equation}\label{api+2}
\Pi^{+2}=\nabla x^{+2}-\Omega^{+2i}x^i-2i\pi^{1+}_q\theta^{1+}_q-2i\pi^{2+}_{\dot q}\theta^{2+}_{\dot q},
\end{equation}
\begin{equation}\label{api-2}
\Pi^{-2}=\nabla x^{-2}-\Omega^{-2i}x^i-2i\pi^{1-}_{\dot q}\theta^{1-}_{\dot q}-2i\pi^{2-}_q\theta^{2-}_q,
\end{equation}
\begin{equation}\label{apii}
\Pi^i=\nabla x^i-\frac12\Omega^{+2i}x^{-2}-\frac12\Omega^{-2i}x^{+2}-i\pi^{1+}_q\gamma^i_{q\dot q}\theta^{1-}_{\dot q}-i\pi^{1-}_{\dot q}\tilde\gamma^i_{\dot qq}\theta^{1+}_{q}+i\pi^{2-}_q\gamma^i_{q\dot q}\theta^{2+}_{\dot q}+i\pi^{2+}_{\dot q}\tilde\gamma^i_{\dot qq}\theta^{2-}_{q},
\end{equation}
\begin{equation}\label{pialfa}
\pi^{1(\underline\alpha)}=\left(\pi^{1+}_q\atop\pi^{1-}_{\dot q}\right)=\left(\nabla\theta^{1+}_q-\frac12\gamma^i_{q\dot r}\theta^{1-}_{\dot r}\Omega^{+2i}\atop  \nabla\theta^{1-}_{\dot q}-\frac12\tilde\gamma^i_{\dot qr}\theta^{1+}_r\Omega^{-2i}\right),
\end{equation}
\begin{equation}\label{pi2-alfa}
\pi^2_{(\underline\alpha)}=\left(\pi^{2-}_q\atop\pi^{2+}_{\dot q}\right)=\left(\nabla\theta^{2-}_q+\frac12\gamma^i_{q\dot r}\theta^{2+}_{\dot r}\Omega^{-2i}\atop \nabla\theta^{2+}_{\dot q}+\frac12\tilde\gamma^i_{\dot qr}\theta^{2-}_r\Omega^{+2i}\right).
\end{equation}
In (\ref{api+2})-(\ref{pi2-alfa}) we introduced covariant differentials $\nabla$ defined as
\begin{equation}\begin{array}{c}
\nabla x^{\pm2}=dx^{\pm2}\pm\Omega^{(0)}x^{\pm2}, \nabla x^i=dx^i-\Omega^{ij}x^j,\\[0.3cm]
\nabla \theta^{\pm}_q=d\theta^{\pm}_q\pm\frac12\Omega^{(0)}\theta^{\pm}_q-\frac14\Omega_{qp}\theta^{\pm}_p, \nabla \theta^{\pm}_{\dot q}=d\theta^{\pm}_{\dot q}\pm\frac12\Omega^{(0)}\theta^{\pm}_{\dot q}-\frac14\Omega_{\dot q\dot p}\theta^{\pm}_{\dot p},
\end{array}\end{equation}
where $\Omega_{qp}\equiv\gamma^{ij}_{qp}\Omega^{ij}$ and $\Omega_{\dot q\dot p}\equiv\tilde\gamma^{ij}_{\dot q\dot p}\Omega^{ij}$.

Analogously, in case of the Type IIB target superspace one has
\begin{equation}\label{bpi+2}
\Pi^{+2}=\nabla x^{+2}-\Omega^{+2i}x^i-2i\pi^{1+}_q\theta^{1+}_q-2i\pi^{2+}_{q}\theta^{2+}_{q},
\end{equation}
\begin{equation}\label{bpi-2}
\Pi^{-2}=\nabla x^{-2}-\Omega^{-2i}x^i-2i\pi^{1-}_{\dot q}\theta^{1-}_{\dot q}-2i\pi^{2-}_{\dot q}\theta^{2-}_{\dot q},
\end{equation}
\begin{equation}\label{bpii}
\Pi^i=\nabla x^i-\frac12\Omega^{+2i}x^{-2}-\frac12\Omega^{-2i}x^{+2}-i\pi^{1+}_q\gamma^i_{q\dot q}\theta^{1-}_{\dot q}-i\pi^{1-}_{\dot q}\tilde\gamma^i_{\dot qq}\theta^{1+}_{q}+(1\leftrightarrow2).
\end{equation}
The definitions of fermionic 1-forms $\pi^{1,2(\underline\alpha)}$ coincide with (\ref{pialfa}). 

The integrability conditions for 1-forms $\Pi^{\underline m}$ and $d\Theta$ are 
\begin{equation}
d\Pi^{\underline m}=-id\Theta^1\sigma^{\underline m}d\Theta^1-id\Theta^2\tilde\sigma^{\underline m}d\Theta^2, dd\Theta^{1\underline\mu}=0, dd\Theta^2_{\underline\mu}=0,\qquad(IIA)
\end{equation}
\begin{equation}
d\Pi^{\underline m}=-id\Theta^1\sigma^{\underline m}d\Theta^1-id\Theta^2\sigma^{\underline m}d\Theta^2, dd\Theta^{1,2\underline\mu}=0,\qquad(IIB)
\end{equation}
that can be considered as Maurer-Cartan equations for the supertranslations in the flat superspace \cite{susy}. In the new string basis defined by (\ref{1.4}), (\ref{1.5}) they transform into
\begin{equation}\label{intbosA}
\nabla\Pi^{(\underline m)}=-i\pi^{1(\underline\alpha)}\sigma^{(\underline m)}_{(\underline\alpha)(\underline\beta)}\pi^{1(\underline\beta)}-i\pi^2_{(\underline\alpha)}\tilde\sigma^{(\underline m)(\underline\alpha)(\underline\beta)}\pi^2_{(\underline\beta)},\qquad(IIA)
\end{equation}
\begin{equation}\label{intbosB}
\nabla\Pi^{(\underline m)}=-i\pi^{1(\underline\alpha)}\sigma^{(\underline m)}_{(\underline\alpha)(\underline\beta)}\pi^{1(\underline\beta)}-(1\leftrightarrow2),\qquad(IIB)
\end{equation}
\begin{equation}\label{intferm}
\nabla\pi^{(\underline\alpha)}=d\pi^{(\underline\alpha)}-\frac14\pi^{(\underline\beta)}\Omega_{(\underline\beta)}{}^{(\underline\alpha)}=0, \nabla\pi_{(\underline\alpha)}=d\pi^{(\underline\alpha)}-\frac14\pi_{(\underline\beta)}\Omega^{(\underline\beta)}{}_{(\underline\alpha)}=0.
\end{equation}

The worldsheet superspace is locally parametrized by the coordinates $z^M=(\xi^m,\eta^{\mu q})$. $\xi^m$ is the $2d$ vector and $\eta^{\mu q}$ is the $2d$ Majorana spinor containing two components with opposite chiralities, index $q=1,...,n$ labels extended worldsheet supersymmetries. The $2d$ superspaces of that kind are referred to as $(n,n)$ superspaces. For the $D=10$ Type II target superspace $n$ can take values from 1 to 8. $n=(8,8)$ Superzweinbein 1-forms in the light-like notations for $SO(1,1)$ group are the following 
\begin{equation}\label{IIA}
e^{\pm2},\quad e^{+}_{q},\quad e^{-}_{q}
\end{equation}
or
\begin{equation}\label{IIB}
e^{\pm2},\quad e^{+}_{q},\quad e^{-}_{\dot q}.
\end{equation}
The first choice is adapted for the description of superworldsheet embedding into the Type IIA target superspace, the second choice is adapted for the Type IIB target superspace. The motivation for this is the structure of irreducible $\kappa-$symmetry transformations of the corresponding component formulations \cite{BZstring}.

In the string basis for the target superspace coordinates (\ref{1.4}), (\ref{1.5}) bosonic embedding equations have the well known form 
\begin{equation}\label{embedb}
\Pi^{+2}=e^{+2},\quad\Pi^{-2}=e^{-2},\quad\Pi^i=0.
\end{equation}
They state that it is always possible by means of proper rotations to equate two bosonic components of the target space supervielbein to the bosonic components of the worldsheet superzweinbein, the rest of the target space supervielbein bosonic components being orthogonal to the string. 
What concerns the fermionic embedding equations there exist two different cases. If $n=8$ one is able to choose
\begin{equation}\label{embedfA}
\pi^{1+}_{q}=e^+_q,\quad\pi^{2-}_{q}=e^-_q
\end{equation}
in the Type IIA case and 
\begin{equation}\label{embedfB}
\pi^{1+}_{q}=e^+_q,\quad\pi^{2-}_{\dot q}=e^-_{\dot q}
\end{equation}
in the Type IIB case. When $n<8$ we are able to introduce additional bosonic $SO(8)$ spinors  $\lambda^a_q(z^M)$, $\lambda^{\dot a}_{\dot q}(z^M)$ in order to adjust covariantly Grassmann components of the target space supervielbein to the Grassmann components of the superzweinbein. For establishing a relation between the NSR string and the Type II GS superstring models the embedding of the $n=(1,1)$ worldsheet superspace is of the primary importance. In this case we have
\begin{equation}\label{embedfA2}
\pi^{1+}_{q}\lambda^1_q=e^+,\quad\pi^{2-}_{q}\lambda^2_q=e^-
\end{equation}
for the Type IIA target superspace and
\begin{equation}\label{embedfB2}
\pi^{1+}_{q}\lambda^1_q=e^+,\quad\pi^{2-}_{\dot q}\lambda^2_{\dot q}=e^-
\end{equation}
in the Type IIB target superspace.

 Target superspace fermionic 1-forms entering (\ref{embedfA2}), (\ref{embedfB2}) have the following general decompositions over the basis of $n=(1,1)$ superzweinbeins
\begin{equation}\label{p1+q}
\pi^{1+}_{q}=e^{+2}\psi^{1+}_{+2q}+e^{-2}\psi^{1+}_{-2q}+e^+\chi^1_q+e^-h^{1+}_{-q},
\end{equation}
\begin{equation}\label{p2-q}
\pi^{2-}_q=e^{+2}\psi^{2-}_{+2q}+e^{-2}\psi^{2-}_{-2q}+e^+h^{2-}_{+q}+e^-\chi^2_q
\end{equation}
in  the Type IIA case, (\ref{p1+q}) and
\begin{equation}\label{p2-dotq}
\pi^{2-}_{\dot q}=e^{+2}\psi^{2-}_{+2\dot q}+e^{-2}\psi^{2-}_{-2\dot q}+e^+h^{2-}_{+\dot q}+e^-\chi^2_{\dot q}
\end{equation}
in the Type IIB case.

Other target superspace fermionic 1-forms decompositions read  
\begin{equation}\label{p1-dotq}
\pi^{1-}_{\dot q}=e^{+2}\psi^{1-}_{+2\dot q}+e^{-2}\psi^{1-}_{-2\dot q}+e^+h^{1-}_{+\dot q}+e^-\chi^1_{\dot q},
\end{equation}
\begin{equation}\label{p2+dotq}
\pi^{2+}_{\dot q}=e^{+2}\psi^{2+}_{+2\dot q}+e^{-2}\psi^{2+}_{-2\dot q}+e^+\chi^2_{\dot q}+e^-h^{2+}_{-\dot q}
\end{equation}
for the Type IIA target superspace and 
\begin{equation}\label{p2+q}
\pi^{2+}_{q}=e^{+2}\psi^{2+}_{+2q}+e^{-2}\psi^{2+}_{-2q}+e^+\chi^2_q+e^-h^{2+}_{-q},
\end{equation}
\begin{equation}\label{p1-dotq'}
\pi^{1-}_{\dot q}=e^{+2}\psi^{1-}_{+2\dot q}+e^{-2}\psi^{1-}_{-2\dot q}+e^+h^{1-}_{+\dot q}+e^-\chi^1_{\dot q}
\end{equation}
for the Type IIB target superspace. In the decompositions (\ref{p1+q})-(\ref{p1-dotq'})  $\psi(z^M)$-superfields are Grassmann ones, other superfields are bosonic.  Thus, relations (\ref{embedfA2}), (\ref{embedfB2}) impose certain restrictions on the decomposition coefficients (\ref{p1+q})-(\ref{p2-dotq}). Namely, from (\ref{embedfA2}) we deduce 
\begin{equation}\label{compA1}
\psi^{1+}_{\pm2q}\lambda^1_q=h^{1+}_{-q}\lambda^1_q=0, \quad\chi^1_q\lambda^1_q=1,
\end{equation}
\begin{equation}\label{compA2}
\psi^{2-}_{\pm2q}\lambda^2_q=h^{2-}_{+q}\lambda^2_q=0,\quad\chi^2_q\lambda^2_q=1
\end{equation}
and from (\ref{embedfB2}) 
\begin{equation}
\psi^{1+}_{\pm2q}\lambda^1_q=h^{1+}_{-q}\lambda^1_q=0, \quad\chi^1_q\lambda^1_q=1,
\end{equation}
\begin{equation}\label{compB2}
\psi^{2-}_{\pm2\dot q}\lambda^2_{\dot q}=h^{2-}_{+\dot q}\lambda^2_{\dot q}=0,\quad\chi^2_{\dot q}\lambda^2_{\dot q}=1.
\end{equation}

Further analysis of embedding lies in the investigation of integrability conditions for the basic superembedding equations in each case together with Maurer-Cartan equations
\begin{equation}
d\Omega^{(\underline m)(\underline n)}-\Omega^{(\underline m)(\underline k)}\Omega_{(\underline k)}{}^{(\underline n)}=\nabla \Omega^{(\underline m)(\underline n)}=0.
\end{equation}
In the case of embedding of strings these equations split as follows
\begin{equation}\label{Gauss}
d\Omega^{(0)}=\frac12\Omega^{+2i}\Omega^{-2i},
\end{equation}
\begin{equation}\label{PC}
d\Omega^{\pm2i}\mp\Omega^{(0)}\Omega^{\pm2i}+\Omega^{ij}\Omega^{\pm2j}=0,
\end{equation}
\begin{equation}\label{Ricci}
d\Omega^{ij}+\Omega^{ik}\Omega^{kj}=-\frac12\Omega^{+2i}\Omega^{-2j}-\frac12\Omega^{-2i}\Omega^{+2j}.
\end{equation}
Equations (\ref{Gauss})-(\ref{Ricci}) are known as Gauss, Codazzi and Ricci equations correspondingly.

In the next section we will examine the embedding of $n=(8,8)$ worldsheet superspace into the $D=10$ Type IIA target superspace. This problem was considered before \cite{BPSTV}, we, however, turn to it again both for the completeness of presentation and for the comparison with the $n=(1,1)$ superworldsheet embedding (Section 4). In Sections 5 and 6 we address the same problems for the target superspace of the Type IIB.

\setcounter{equation}{0}

\section{Embedding of the $n=(8,8)$ superworldsheet into $D=~10$ $N=2A$ target superspace}

This case is described by superembedding equations (\ref{embedb}), (\ref{embedfA}). We begin with deriving their integrability conditions. Consider first the differentials of bosonic embedding equations (\ref{embedb})
\begin{equation}\label{dembedb}
d\Pi^{+2}=de^{+2},\quad d\Pi^{-2}=de^{-2},\quad d\Pi^i=0.
\end{equation}
Splitting (\ref{intbosA}) into the tangent and orthogonal to the worldsheet parts one deduces in general case that 
\begin{equation}\label{dpi+2a}
d\Pi^{+2}=\Omega^{(0)}\Pi^{+2}-\Omega^{+2i}\Pi^i-2i\pi^{1+}_q\pi^{1+}_q-2i\pi^{2+}_{\dot q}\pi^{2+}_{\dot q},
\end{equation}
\begin{equation}
d\Pi^{-2}=-\Omega^{(0)}\Pi^{-2}-\Omega^{-2i}\Pi^i-2i\pi^{1-}_{\dot q}\pi^{1-}_{\dot q}-2i\pi^{2-}_{q}\pi^{2-}_{q},
\end{equation}
\begin{equation}\label{dpiia}
d\Pi^i=-\frac12\Omega^{+2i}\Pi^{-2}-\frac12\Omega^{-2i}\Pi^{+2}-\Omega^{ij}\Pi^j-2i\pi^{1+}_q\gamma^i_{q\dot q}\pi^{1-}_{\dot q}+2i\pi^{2-}_q\gamma^i_{q\dot q}\pi^{2+}_{\dot q}.
\end{equation}
Torsion 2-forms for the worldsheet superspace under consideration are defined as
\begin{equation}\label{tors}
T^{\pm2}=de^{\pm2}\mp\omega^{(0)}e^{\pm2},
\end{equation}
\begin{equation}\label{torq}
T^{\pm}_{ q}=de^{\pm}_{ q}\mp\frac12\omega^{(0)}e^{\pm}_{ q}+\frac14\omega_{qp}e^{\pm}_{ p},
\end{equation}
where $\omega^{(0)}$ and $\omega^{qp}$ are intrinsic connection 1-forms. Substituting equations (\ref{dpi+2a})-(\ref{torq}) back into (\ref{dembedb}) and taking into account (\ref{embedb}) one finds
\begin{equation}
\Omega^{(0)}e^{+2}-2ie^{1+}_q e^{1+}_q-2i\pi^{2+}_{\dot q}\pi^{2+}_{\dot q}=T^{+2}+\omega^{(0)}e^{+2},
\end{equation}
\begin{equation}
-\Omega^{(0)}e^{-2}-2i\pi^{1-}_{\dot q}\pi^{1-}_{\dot q}-2ie^{2-}_{q}e^{2-}_{q}=T^{-2}-\omega^{(0)}e^{+2},
\end{equation}
\begin{equation}\label{dpii8a}
-\frac12\Omega^{+2i}e^{-2}-\frac12\Omega^{-2i}e^{+2}-2ie^{1+}_q\gamma^i_{q\dot q}\pi^{1-}_{\dot q}+2ie^{2-}_q\gamma^i_{q\dot q}\pi^{2+}_{\dot q}=0.
\end{equation}
It is possible to identify induced and intrinsic $SO(1,1)$ connections by setting
\begin{equation}
\Omega^{(0)}=\omega^{(0)}.
\end{equation}
(Another possible way of consideration is to impose supergravity constraints on the intrinsic torsion 2-form and to introduce modified connections $\Omega(D)=\Omega(d)-\omega(d)$. As was shown in \cite{BPSTV} in case of strings $\Omega(D)=0$, so induced and intrinsic connections coincide.)
Thus, $+2$ and $-2$ torsion components acquire the form
\begin{equation}\label{t+-28a}
T^{+2}=-2ie^{+}_{q}e^{+}_{q}-2i\pi^{2+}_{\dot q}\pi^{2+}_{\dot q},\quad T^{-2}=-2i\pi^{1-}_{\dot q}\pi^{1-}_{\dot q}-2ie^{-}_{q}e^{-}_{q}.
\end{equation}
Application of differential operation to the fermionic embedding equations (\ref{embedfA}) and the utilization of (\ref{torq}) together with (\ref{intferm}) 
\begin{equation}\label{dp1+q}
d\pi^{1+}_{q}=\frac12\Omega^{(0)}\pi^{1+}_{q}-\frac14\Omega_{qp}\pi^{1+}_p-\frac12\gamma^i_{q\dot p}\Omega^{+2i}\pi^{1-}_{\dot p},
\end{equation}
\begin{equation}\label{dp2-q}
d\pi^{2-}_{q}=-\frac12\Omega^{(0)}\pi^{2-}_{q}-\frac14\Omega_{qp}\pi^{2-}_p+\frac12\gamma^i_{q\dot p}\Omega^{-2i}\pi^{2+}_{\dot p},
\end{equation}
leads  to 
\begin{equation}
-\frac14\Omega_{qp}e^{+}_p-\frac12\gamma^i_{q\dot p}\Omega^{+2i}\pi^{1-}_{\dot p}=T^{+}_{q}-\frac14\omega_{qp}e^{+}_p,
\end{equation}
\begin{equation}
-\frac14\Omega_{qp}e^{-}_p+\frac12\gamma^i_{q\dot p}\Omega^{-2i}\pi^{2+}_{\dot p}=T^{-}_{q}-\frac14\omega_{qp}e^-_p.
\end{equation}
We are able to identify induced and intrinsic $SO(8)$ connection 1-forms
\begin{equation}
\Omega_{qp}=\omega_{qp}
\end{equation}
to obtain in this way
\begin{equation}\label{tferm8a}
T^+_q=-\frac12\gamma^i_{q\dot p}\Omega^{+2i}\pi^{1-}_{\dot p},\quad T^-_q=\frac12\gamma^i_{q\dot p}\Omega^{-2i}\pi^{2+}_{\dot p}.
\end{equation}
Then we should explore the integrability conditions for those target space supervielbein fermionic components that were not identified with the superzweinbein fermionic components, i.e. for $\pi^{1-}_{\dot q}$, $\pi^{2+}_{\dot q}$. On one hand, we have from (\ref{intferm}) 
\begin{equation}\label{dp1-dotq}
d\pi^{1-}_{\dot q}=-\frac12\Omega^{(0)}\pi^{1-}_{\dot q}-\frac14\Omega_{\dot q\dot p}\pi^{1-}_{\dot p}-\frac12\tilde\gamma^i_{\dot qp}\Omega^{-2i}\pi^{1+}_p,
\end{equation}
\begin{equation}\label{dp2+dotq}
d\pi^{2+}_{\dot q}=\frac12\Omega^{(0)}\pi^{2+}_{\dot q}-\frac14\Omega_{\dot q\dot p}\pi^{2+}_{\dot p}+\frac12\tilde\gamma^i_{\dot qp}\Omega^{+2i}\pi^{2-}_p,
\end{equation}
but on the other hand, using the decompositions of $\pi^{1-}_{\dot q}$ and $\pi^{2+}_{\dot q}$ over the superworldsheet basis  
\begin{equation}\begin{array}{c}\label{decomp8a}
\pi^{1-}_{\dot q}=e^{+2}\psi^{1-}_{+2\dot q}+e^{-2}\psi^{1-}_{-2\dot q}+e^+_qh^{1-}_{+q\dot q}+e^-_q\chi^1_{q\dot q},\\[0.3cm] \pi^{2+}_{\dot q}=e^{+2}\psi^{2+}_{+2\dot q}+e^{-2}\psi^{2+}_{-2\dot q}+e^+_q\chi^{2}_{q\dot q}+e^-_qh^{2+}_{-q\dot q} 
\end{array}\end{equation}
we derive
\begin{equation}\begin{array}{l}
e^{+2}\nabla \psi^{1-}_{+2\dot q}+e^{-2}\nabla \psi^{1-}_{-2\dot q}+e^+_p\nabla h^{1-}_{+p\dot q}+e^-_p\nabla\chi^1_{p\dot q}+T^{+2}\psi^{1-}_{+2\dot q}+T^{-2}\psi^{1-}_{-2\dot q}+T^+_{p}h^{1-}_{+p\dot q}+T^-_p\chi^1_{p\dot q}\\[0.3cm]
=\frac12e^+_p\gamma^i_{p\dot q}\Omega^{-2i},\label{56}
\end{array}\end{equation}
\begin{equation}\begin{array}{l}
e^{+2}\nabla \psi^{2+}_{+2\dot q}+e^{-2}\nabla \psi^{2+}_{-2\dot q}+e^+_p\nabla\chi^2_{p\dot q}+e^-_p\nabla h^{2+}_{-p\dot q}+T^{+2}\psi^{2+}_{+2\dot q}+T^{-2}\psi^{2+}_{-2\dot q}+T^+_p\chi^2_{p\dot q}+T^-_{p}h^{2+}_{-p\dot q}\\[0.3cm]
=-\frac12e^-_p\gamma^i_{p\dot q}\Omega^{+2i}.\label{57}
\end{array}\end{equation}

So, we have found all the integrability conditions for the embedding under consideration and turn now to their analysis. Consider, first, equation  (\ref{dpii8a}). Its spinor-spinor components read
\begin{equation}\label{e+e+}
\gamma^i_{q\dot q}h^{1-}_{+p\dot q}+\gamma^i_{p\dot q}h^{1-}_{+q\dot q}=0,\ 
\gamma^i_{q\dot q}h^{2+}_{-p\dot q}+\gamma^i_{p\dot q}h^{2+}_{-q\dot q}=0,\ 
-\gamma^i_{q\dot q}\chi^1_{\dot qp}+\gamma^i_{p\dot q}\chi^2_{\dot qq}=0.
\end{equation}
To analyse equations (\ref{e+e+}) it is necessary to decompose $h^{1-}_{+q\dot q}$, $h^{2+}_{-q\dot q}$, $\chi^{1,2}_{q\dot q}$ on the $SO(8)$ irreducible parts
\begin{equation}\label{1+3}
h^{1-}_{+q\dot q}=\gamma^i_{q\dot q}h^{1-i}_++\gamma^{ijk}_{q\dot q}h^{1-ijk}_+,\ h^{2+}_{-q\dot q}=\gamma^i_{q\dot q}h^{2+i}_-+\gamma^{ijk}_{q\dot q}h^{2+ijk}_-,\ \chi^{1,2}_{q\dot q}=\gamma^i_{q\dot q}\chi^{1,2i}+\gamma^{ijk}_{q\dot q}\chi^{1,2ijk}.
\end{equation}
Substituting (\ref{1+3}) back into (\ref{e+e+}) after some manipulations with $SO(8)$ $\sigma-$matrices one obtains
\begin{equation}\label{e+qe-pmod}
h^{1-}_{+q\dot q}=h^{2+}_{-q\dot q}=\chi^{1,2i}=0,\ \chi^{1ijk}=-\chi^{2ijk}.
\end{equation}
Then using Bianchi identities for the induced supertorsion it is possible to show that 
\begin{equation}\label{e+qe-pmod'}
\chi^{1ijk}=0.
\end{equation}
  Other components of (\ref{dpii8a}) define the second fundamental form through the embedding functions
\begin{equation}\label{secform1}
\Omega^{+2i}_{+2}-\Omega^{-2i}_{-2}=0,
\end{equation}
\begin{equation}
\frac12\Omega^{-2i}_{+q}+2i\gamma^i_{q\dot q}\psi^{1-}_{+2\dot q}=0,\  -\frac12\Omega^{-2i}_{-q}+2i\gamma^i_{q\dot q}\psi^{2+}_{+2\dot q}=0,
\end{equation}
\begin{equation}\label{secform3}
\frac12\Omega^{+2i}_{+q}+2i\gamma^i_{q\dot q}\psi^{1-}_{-2\dot q}=0,\  -\frac12\Omega^{+2i}_{-q}+2i\gamma^i_{q\dot q}\psi^{2+}_{-2\dot q}=0.
\end{equation}
Nonzero induced torsion components that remain upon utilization  of (\ref{e+qe-pmod}), (\ref{e+qe-pmod'}) are
\begin{equation}\label{t+2+2-2}
T^{+2}_{+2-2}=4i\psi^{2+}_{+2\dot q}\psi^{2+}_{-2\dot q},\quad T^{+2}_{+q+p}=-4i\delta_{qp},
\end{equation}
\begin{equation}
T^{-2}_{+2-2}=4i\psi^{1-}_{+2\dot q}\psi^{1-}_{-2\dot q},\quad T^{+2}_{+q+p}=-4i\delta_{qp},
\end{equation}
and 
\begin{equation}\label{t+q+2-28a}
T^{+q}_{+2-2}=\frac12\gamma^i_{q\dot q}\Omega^{+2i}_{+2}\psi^{1-}_{-2\dot q}-\frac12\gamma^i_{q\dot q}\Omega^{+2i}_{-2}\psi^{1-}_{+2\dot q},
\end{equation}
\begin{equation}\label{t+qpb8a}
T^{+q}_{+p\pm2}=\frac12\gamma^i_{q\dot q}\Omega^{+2i}_{+p}\psi^{1-}_{\pm2\dot q},\quad  T^{+q}_{-p\pm2}=\frac12\gamma^i_{q\dot q}\Omega^{+2i}_{-p}\psi^{1-}_{\pm2\dot q};
\end{equation}
\begin{equation}\label{t-q+2-28a}
T^{-q}_{+2-2}=\frac12\gamma^i_{q\dot q}\Omega^{-2i}_{-2}\psi^{2+}_{+2\dot q}-\frac12\gamma^i_{q\dot q}\Omega^{-2i}_{+2}\psi^{1-}_{-2\dot q},
\end{equation}
\begin{equation}\label{t-qpb8a}
T^{-q}_{+p\pm2}=-\frac12\gamma^i_{q\dot q}\Omega^{-2i}_{+p}\psi^{2+}_{\pm2\dot q},\quad  T^{-q}_{-p\pm2}=\frac12\gamma^i_{q\dot q}\Omega^{-2i}_{-p}\psi^{2+}_{\pm2\dot q}.
\end{equation}

We are in a position to consider the consequences of the integrability conditions (\ref{56}), (\ref{57}). From (\ref{57}) after taking into account (\ref{e+qe-pmod})-(\ref{t-qpb8a}) we deduce
\begin{equation}\label{57'}
\Omega^{+2i}_{+q}=0 \Rightarrow \psi^{1-}_{-2\dot q}=0,\  \psi^{2+}_{+2\dot q}=0\Rightarrow \Omega^{-2i}_{-q}=0,\  \psi^{2+}_{-2\dot q}=-\frac{i}{32}\tilde\gamma^i_{\dot qq}\Omega^{+2i}_{-q}, 
\end{equation}
\begin{equation}\label{57''}
\Omega^{+2i}_{+2}=\Omega^{-2i}_{-2}=0,\quad\Omega^{+2i}_{-2}=\frac14\nabla_{-p}\gamma^i_{p\dot q}\psi^{2+}_{-2\dot q},
\end{equation}
\begin{equation}
\nabla_{+p}\psi^{2+}_{-2\dot q}=0,\quad\nabla_{+2}\psi^{2+}_{-2\dot q}=0.
\end{equation}
 From the integrability conditions for $\pi^{1-}_{\dot q}$ (\ref{56}) there follow the relations
\begin{equation}
\psi^{1-}_{+2\dot q}=\frac{i}{32}\tilde\gamma^i_{\dot qp}\Omega^{-2i}_{+p},\ \Omega^{-2i}_{+2}=-\frac14\nabla_{+q}\gamma^i_{q\dot q}\psi^{1-}_{+2\dot q},
\end{equation}
\begin{equation}\label{56'}
\nabla_{-q}\psi^{1-}_{+2\dot q}=0,\ \nabla_{-2}\psi^{1-}_{+2\dot q}=0.
\end{equation}
We observe that from the variety of functions entering (\ref{decomp8a}) there remained only two chiral superfields $\psi^{1-}_{+2\dot q}(z^M)$ and $\psi^{2+}_{-2\dot q}(z^M)$. 

Maurer-Cartan equations (\ref{Gauss}), (\ref{PC}), (\ref{Ricci}) in the considered case do not contain any additional dynamical information.

The nonzero second fundamental form components are expressed through $\psi$ superfileds and their derivatives
\begin{equation}
\Omega^{-2i}_{+q}=-4i\gamma^i_{q\dot q}\psi^{1-}_{+2\dot q},\quad\Omega^{+2i}_{-q}=4i\gamma^i_{q\dot q}\psi^{2+}_{-2\dot q},
\end{equation}
\begin{equation}
\Omega^{+2i}_{-2}=\frac14\nabla_{-p}\gamma^i_{p\dot q}\psi^{2+}_{-2\dot q},\quad\Omega^{-2i}_{+2}=-\frac14\nabla_{+p}\gamma^i_{p\dot q}\psi^{1-}_{+2\dot q}.
\end{equation}
Supertorsion components (\ref{t+2+2-2})-(\ref{t-qpb8a}) after substitution of (\ref{57'})-(\ref{56'}) reduce to 
\begin{equation}
T^{\pm2}_{\pm q\pm p}=-4i\delta_{qp},
\end{equation}
\begin{equation}
T^{+q}_{-p+2}=-2i\gamma^i_{q\dot q}\psi^{1-}_{+2\dot q}\gamma^i_{p\dot p}\psi^{2+}_{-2\dot q},\quad T^{+q}_{+2-2}=\frac{i}{16}\nabla_{-p}T^{+q}_{-p+2},
\end{equation}
\begin{equation}
T^{-q}_{+p-2}=-2i\gamma^i_{q\dot q}\psi^{2+}_{-2\dot q}\gamma^i_{p\dot p}\psi^{1-}_{+2\dot q},\quad T^{+q}_{+2-2}=-\frac{i}{16}\nabla_{+p}T^{-q}_{+p-2}.
\end{equation}

Explicit expressions for the original target superspace supervielbeins follow from (\ref{ba})-(\ref{fa})
\begin{equation}\label{72}
d\Theta^{1\underline\mu}=v^{-\underline\mu}_q\pi^{1+}_q+v^{+\underline\mu}_{\dot q}\pi^{1-}_{\dot q}=e^{+2}\psi^{1-}_{+2\dot q}v^{\underline\mu+}_{\dot q}+e^+_qv^{\underline\mu-}_q,
\end{equation}
\begin{equation}\label{73}
d\Theta^2_{\underline\mu}=v^{+}_{\underline\mu q}\pi^{2-}_q+v^{-}_{\underline\mu\dot q}\pi^{2+}_{\dot q},
=e^{-2}\psi^{2+}_{-2\dot q}v^-_{\underline\mu\dot q}+e^-_qv^{+}_{\underline\mu q},
\end{equation}
\begin{equation}
\Pi^{\underline m}=u^{\underline m+2}e^{-2}+u^{\underline m-2}e^{+2}.
\end{equation}

Let us show that the considered embedding is on-shell. The Type IIA superstring equations of motion in the twistor-like Lorentz harmonic formulation read
\begin{equation}\label{sfg2af}
D_{-2}\theta^{1\underline\mu}(\xi^m)v^-_{\underline\mu\dot q}=0,\quad D_{+2}\theta^2_{\underline\mu}(\xi^m)v_{\dot q}^{+\underline\mu}=0,
\end{equation}
\begin{equation}\label{sfg2ab}
\frac12\partial_{\mu}\left(e\sum\limits_{\pm}e^{\mu\pm2}u^{\mp2}_{\underline m}\right)-i\epsilon^{\mu\nu}\left(\partial_{\mu}\theta^{1\underline\mu}(\xi^m)\sigma_{\underline m\underline{\mu\nu}}\partial_{\nu}\theta^{1\underline\nu}(\xi^m)-\partial_{\mu}\theta^2_{\underline\mu}(\xi^m)\tilde\sigma_{\underline m}^{\underline{\mu\nu}}\partial_{\nu}\theta^2_{\underline\nu}(\xi^m)\right)=0.
\end{equation}
Upon utilization of (\ref{cartan}), (\ref{1.6}), (\ref{fa}) these equations can be written in terms of the second fundamental form and functions $\psi^{1+}_{\pm2q}(\xi^m)\!=\! D_{\pm2}\theta^{1\underline\mu}v^+_{\underline\mu q}$, $\psi^{1-}_{\pm2\dot q}(\xi^m)\!=\! D_{\pm2}\theta^{1\underline\mu}v^-_{\underline\mu\dot q}$, $\psi^{2+}_{\pm2\dot q}(\xi^m)\!=\! D_{\pm2}\theta^2_{\underline\mu}v^{+\underline\mu}_{\dot q}$, $\psi^{2-}_{\pm2 q}(\xi^m)\!=\! D_{\pm2}\theta^2_{\underline\mu}v^{-\underline\mu}_{q}$
\begin{equation}\label{2.51}
\psi^{1-}_{-2\dot q}=\psi^{2+}_{+2\dot q}=0,
\end{equation}
\begin{equation}\label{2.52}
\Omega^{+2i}_{+2}+\Omega^{-2i}_{-2}+4i\left(\psi^{1+}_{+2q}\gamma^i_{q\dot q}\psi^{1-}_{-2\dot q}-\psi^{1+}_{-2q}\gamma^i_{q\dot q}\psi^{1-}_{+2\dot q}+\psi^{2-}_{+2q}\gamma^i_{q\dot q}\psi^{2+}_{-2\dot q}-\psi^{2-}_{-2q}\gamma^i_{q\dot q}\psi^{2+}_{+2\dot q}\right)=0.
\end{equation}
When deriving bosonic equation (\ref{2.52}) we have taken into account that the tangent to the worldsheet part of (\ref{sfg2ab}) does not contain independent equations due to the reparametrization symmetry. Superstring equations of motion in the form (\ref{2.51}), (\ref{2.52}) permit natural worldsheet superfield generalization. Then noting that from the fermionic embedding equations (\ref{embedfA}) it follows that $\psi^{1+}_{\pm2q}(z^M)\!=\!\psi^{2-}_{\pm2q}(z^M)\!=\!0$ we find that  (\ref{2.52}) acquires the form 
\begin{equation}\label{3.53}
\Omega^{+2i}_{+2}(z^M)+\Omega^{-2i}_{-2}(z^M)=0.
\end{equation}
It is nothing but the minimal embedding condition known from the bosonic theory. On the other hand, (\ref{3.53}) is satisfied by equations (\ref{57''}) that stem from the integrability conditions for superembedding equations.

\setcounter{equation}{0}

\section{Embedding of the $n=(1,1)$ superworldsheet into $D=~10$ $N=2A$ target superspace}

Following the guidelines of the previous Section first we have to derive intergrability conditions for the basic superembedding equations (\ref{embedb}), (\ref{embedfA2}). Applying external differential to (\ref{embedb}), and using (\ref{dpi+2a})-(\ref{dpiia}), and  the definition of the supertorsion 2-forms for the $n=(1,1)$ worldsheet superspace
\begin{equation}\label{torpm2}
T^{\pm2}=de^{\pm2}\mp\omega^{(0)}e^{\pm2},
\end{equation}
\begin{equation}\label{torpm}
T^{\pm}=de^{\pm }\mp\frac12\omega^{(0)}e^{\pm }
\end{equation}
we derive
\begin{equation}\label{dpi+2a1}
\Omega^{(0)}e^{+2}-2i\pi^{1+}_q\pi^{1+}_q-2i\pi^{2+}_{\dot q}\pi^{2+}_{\dot q}=T^{+2}+\omega^{(0)}e^{+2},
\end{equation}
\begin{equation}
-\Omega^{(0)}e^{-2}-2i\pi^{1-}_{\dot q}\pi^{1-}_{\dot q}-2i\pi^{2-}_{q}\pi^{2-}_{q}=T^{-2}-\omega^{(0)}e^{-2},
\end{equation}
\begin{equation}\label{dpiia1}
-\frac12\Omega^{+2i}e^{-2}-\frac12\Omega^{-2i}e^{+2}-2i\pi^{1+}_q\gamma^i_{q\dot q}\pi^{1-}_{\dot q}+2i\pi^{2-}_q\gamma^i_{q\dot q}\pi^{2+}_{\dot q}=0.
\end{equation}
Like in Section 2 we identify $SO(1,1)$ induced and intrinsic connections 
\begin{equation}\label{79}
\Omega^{(0)}=\omega^{(0)},
\end{equation}
and thus obtain the expressions for the $\pm2$ components of the induced supertorsion 
\begin{equation}\label{t+-21a}
T^{+2}=-2i\pi^{1+}_{q}\pi^{1+}_{q}-2i\pi^{2+}_{\dot q}\pi^{2+}_{\dot q},\quad T^{-2}=-2i\pi^{1-}_{\dot q}\pi^{1-}_{\dot q}-2i\pi^{2-}_{q}\pi^{2-}_{q}.
\end{equation}

Differentiating of (\ref{embedfA2}) and using (\ref{embedfA2}), (\ref{79}) yields
\begin{equation}\label{t+A1}
T^+=\pi^{1+}_{q}\nabla\lambda^1_q-\frac12\lambda^1_q\gamma^i_{q\dot q}\Omega^{+2i}\pi^{1-}_{\dot q},
\end{equation}
\begin{equation}\label{t-A1}
T^-=\pi^{2-}_{q}\nabla\lambda^2_q+\frac12\lambda^2_q\gamma^i_{q\dot q}\Omega^{-2i}\pi^{2+}_{\dot q},
\end{equation}
where we have introduced the covariant differential for the auxiliary spinors $\lambda^{1,2}_q(z^M)$ as 
\begin{equation}
\nabla\lambda^{1,2}_q=d\lambda^{1,2}_q-\frac14\Omega_{qp}\lambda^{1,2}_p.
\end{equation}
The integrability conditions for $\pi^{1+}_q, \pi^{2-}_q$ and $\pi^{1-}_{\dot q}, \pi^{2+}_{\dot q}$ components of the target space supervielbein also should be taken into account. The substitution of general expressions (\ref{p1+q}), (\ref{p2-q}), (\ref{p1-dotq}), (\ref{p2+dotq}) into (\ref{dp1+q}), (\ref{dp2-q}), (\ref{dp1-dotq}), (\ref{dp2+dotq}) gives
\begin{equation}\label{int1+q}\begin{array}{l}
e^{+2}\nabla\psi^{1+}_{+2q}+e^{-2}\nabla\psi^{1+}_{-2q}+e^+\nabla\chi^1_q+e^-\nabla h^{1+}_{-q}+T^{+2}\psi^{1+}_{+2q}+T^{-2}\psi^{1+}_{-2q}+T^+\chi^1_q+T^-h^{1+}_{-q}\\[0.3cm]
=-\frac12\gamma^i_{q\dot p}\Omega^{+2i}\pi^{1-}_{\dot p},
\end{array}\end{equation}
\begin{equation}\label{int2-q}\begin{array}{l}
e^{+2}\nabla\psi^{2-}_{+2q}+e^{-2}\nabla\psi^{2-}_{-2q}+e^+\nabla h^{2-}_{+q}+e^-\nabla\chi^2_q+T^{+2}\psi^{2-}_{+2q}+T^{-2}\psi^{2-}_{-2q}+T^+h^{2-}_{+q}+T^-\chi^2_q\\[0.3cm]
=\frac12\gamma^i_{q\dot p}\Omega^{-2i}\pi^{2+}_{\dot p},
\end{array}\end{equation}
\begin{equation}\label{int1-dotq1a}\begin{array}{l}
e^{+2}\nabla\psi^{1-}_{+2\dot q}+e^{-2}\nabla\psi^{1-}_{-2\dot q}+e^+\nabla h^{1-}_{+\dot q}+e^-\nabla\chi^1_{\dot q}+T^{+2}\psi^{1-}_{+2\dot q}+T^{-2}\psi^{1-}_{-2\dot q}+T^+h^{1-}_{+\dot q}+T^-\chi^1_{\dot q}\\[0.3cm]
=-\frac12\tilde\gamma^i_{\dot qp}\Omega^{-2i}\pi^{1+}_p,
\end{array}\end{equation}
\begin{equation}\label{int2+dotq1a}\begin{array}{l}
e^{+2}\nabla\psi^{2+}_{+2\dot q}+e^{-2}\nabla\psi^{2+}_{-2\dot q}+e^+\nabla\chi^{2}_{\dot q}+e^-\nabla h^{2+}_{-\dot q}+T^{+2}\psi^{2+}_{+2\dot q}+T^{-2}\psi^{2+}_{-2\dot q}+T^+\chi^{2}_{\dot q}+T^- h^{2+}_{-\dot q}\\[0.3cm]
=\frac12\tilde\gamma^i_{\dot qp}\Omega^{+2i}\pi^{2-}_p.
\end{array}\end{equation}
Note, that the multiplication of (\ref{int1+q}) by $\lambda^1_q$ and of (\ref{int2-q}) by $\lambda^2_q$ produces Eqs.(\ref{t+A1}), (\ref{t-A1}) for the induced supertorsion $+$ and $-$ components provided the integrability conditions for (\ref{compA1}), (\ref{compA2}) are used. 

The expressions for the  induced worldsheet supertorsion (\ref{t+-21a})-(\ref{t-A1}) do not yield $2d$ $n=(1,1)$ supergravity constraints in contrast with the $n=(8,8)$ case, so let us contemplate the consequences of their imposition. Conventional $2d$ $n=(1,1)$ supergravity constraints have the form \cite{Howe}
\begin{equation}\label{2dsugra}
T^{\pm2}_{\pm\pm}=-4i,\ T^{+2}_{\pm-}=T^{-2}_{\pm+}=0,\ T^{\pm2}_{+2-2}=0,\ T^{\pm}_{\pm\pm}=T^{+}_{\pm-}=T^-_{\pm+}=0.
\end{equation}
The consideration of Bianchi identities allows to represent all nonzero supertorsion components through the single scalar superfield $S(z^M)$ and its derivatives
\begin{equation}\label{t0}
T^{+2}_{\pm2-}=T^{+2}_{\pm2+}=T^{-2}_{\pm2-}=T^{-2}_{\pm2+}=0, T^{+}_{\pm\pm2}=T^+_{+-2}=T^-_{\pm\pm2}=T^-_{-+2}=0,
\end{equation}
\begin{equation}\label{t-s}
T^+_{-+2}=S,\ T^{-}_{+-2}=-S,\ T^{\pm}_{+2-2}=\mp\frac{i}{2}\nabla_{\mp}S.
\end{equation}
Below we will adduce explicit expression for $S$ in terms of the second fundamental form. 

The insertion of  (\ref{2dsugra})-(\ref{t-s}) into (\ref{t+-21a})-(\ref{t-A1}) results in the following algebraic equations
\begin{equation}\label{h0}
h^{1+}_{-q}=h^{2-}_{+q}=0,\ h^{1-}_{+\dot q}=h^{2+}_{-\dot q}=0,
\end{equation}
\begin{equation}\label{norma}
\chi^1_q\chi^1_q+\chi^2_{\dot q}\chi^2_{\dot q}=1,\ \chi^2_q\chi^2_q+\chi^1_{\dot q}\chi^1_{\dot q}=1.
\end{equation}

From (\ref{dpiia1}) in view of (\ref{h0}), (\ref{norma}) it follows that 
\begin{equation}\label{pi1a+-}
\chi^1_q\gamma^i_{q\dot q}\chi^1_{\dot q}-\chi^2_q\gamma^i_{q\dot q}\chi^2_{\dot q}=0,
\end{equation}
\begin{equation}\label{pi1a+-cons}
\Omega^{-2i}_{-}=-4i\psi^{1+}_{+2q}\gamma^i_{q\dot q}\chi^1_{\dot q}+4i\chi^2_q\gamma^i_{q\dot q}\psi^{2+}_{+2\dot q},\quad \Omega^{+2i}_+=-4i\chi^1_q\gamma^i_{q\dot q}\psi^{1-}_{-2\dot q}+4i\psi^{2-}_{-2q}\gamma^i_{q\dot q}\chi^2_{\dot q},
\end{equation}
\begin{equation}
\Omega^{+2i}_-=-4i\psi^{1+}_{-2q}\gamma^i_{q\dot q}\chi^1_{\dot q}+4i\chi^2_q\gamma^i_{q\dot q}\psi^{2+}_{-2\dot q},\quad  \Omega^{-2i}_+=-4i\chi^{1}_{q}\gamma^i_{q\dot q}\psi^{1-}_{+2\dot q}+4i\psi^{2-}_{+2q}\gamma^i_{q\dot q}\chi^{2}_{\dot q}.
\end{equation}
\begin{equation}\label{pi1a+2-2}
\Omega^{-2i}_{-2}-\Omega^{+2i}_{+2}=4i\psi^{1+}_{+2q}\gamma^i_{q\dot q}\psi^{1-}_{-2\dot q}-4i\psi^{1+}_{-2q}\gamma^i_{q\dot q}\psi^{1-}_{+2\dot q}-4i\psi^{2-}_{+2q}\gamma^i_{q\dot q}\psi^{2+}_{-2\dot q}+4i\psi^{2-}_{-2q}\gamma^i_{q\dot q}\psi^{2+}_{+2\dot q}.
\end{equation}

From the integrability conditions (\ref{int1+q}) one infers as the spinor-spinor component equations 
\begin{equation}\label{int1+q1} 
\psi^{1+}_{+2q}=-\frac{i}{2}\nabla_{+}\chi^1_q,\ \psi^{1+}_{-2q}=\frac{i}{4}\Omega^{+2i}_-\gamma^i_{q\dot q}\chi ^1_{\dot q},
\end{equation} 
\begin{equation}\label{int1+q2} \nabla_-\chi^1_q=\frac12\Omega^{+2i}_+\gamma^i_{q\dot q}\chi^1_{\dot q}.
\end{equation} 
Analogously Eq.(\ref{int2-q}) yields that 
\begin{equation}\label{int2-q1}
\psi^{2-}_{-2q}=-\frac{i}{2}\nabla_{-}\chi^2_q,\ \psi^{2-}_{+2q}=-\frac{i}{4}\Omega^{-2i}_+\gamma^i_{q\dot q}\chi ^2_{\dot q},
\end{equation}
\begin{equation}\label{int2-q2}
\nabla_+\chi^2_q=-\frac12\Omega^{-2i}_-\gamma^i_{q\dot q}\chi^2_{\dot q}.
\end{equation}
Finally equations
\begin{equation}\label{int1-dotq1}
\psi^{1-}_{-2\dot q}=-\frac{i}{2}\nabla_{-}\chi^1_{\dot q},\ \psi^{1-}_{+2\dot q}=\frac{i}{4}\Omega^{-2i}_+\tilde\gamma^i_{\dot qq}\chi ^1_{q},
\end{equation}
\begin{equation}\label{int1-dotq2}
\nabla_+\chi^1_{\dot q}=\frac12\Omega^{-2i}_-\tilde\gamma^i_{\dot qq}\chi^1_q,
\end{equation}
\begin{equation}\label{int2+dotq1}
\psi^{2+}_{+2\dot q}=-\frac{i}{2}\nabla_{+}\chi^2_{\dot q},\ \psi^{2+}_{-2\dot q}=-\frac{i}{4}\Omega^{+2i}_-\tilde\gamma^i_{\dot qq}\chi ^2_{q},
\end{equation}
\begin{equation}\label{int2+dotq2}
\nabla_-\chi^2_{\dot q}=-\frac12\Omega^{+2i}_+\tilde\gamma^i_{\dot qq}\chi^2_q
\end{equation}
emanate from (\ref{int1-dotq1a}) and (\ref{int2+dotq1a}). Spinor-vector and vector-vector component equations of (\ref{int1+q})-(\ref{int2+dotq1a}) are the consequences of the adduced ones. Then the scalar superfield $S$ that enters expressions for the induced supertorsion can be presented as 
\begin{equation}
S=\frac{i}{8}\left(\Omega^{+2i}_+\Omega^{-2i}_-+\Omega^{+2i}_-\Omega^{-2i}_+\right).
\end{equation}

Similarly to the $D=3$ $N=2$ target superspace case \cite{Galperin} considered superembedding is off-shell. This can be easily seen from Eqs. (\ref{int1-dotq1}), (\ref{int2+dotq1})
\begin{equation}
\psi^{1-}_{-2\dot q}=-\frac{i}{2}\nabla_{-}\chi^1_{\dot q},\ \psi^{2+}_{+2\dot q}=-\frac{i}{2}\nabla_{+}\chi^2_{\dot q}
\end{equation}
that have nonzero r.h.s. compared to the Type IIA superstring fermionic equations of motion (\ref{2.51}).

It is possible to consider the case when the auxiliary spinor superfields $\lambda^{1,2}_q(z^M)$ are identified with the superfields $\chi^{1,2}_q(z^M)$ originating from the decompositions (\ref{p1+q}), (\ref{p2-q})
\begin{equation} \lambda^1_q=\chi^1_q,\  \lambda^2_q=\chi^2_q.
\end{equation} 
This scheme can be regarded as the "minimal" one in the sence that it involves only the superfields originating from the decompositions of the target space supervielbein over the superzweinbein basis. From (\ref{compA1}), (\ref{compA2}) we deduce that $\lambda^{1,2}_q$ possess the unit norm. Let us examine the situation when in addition the worldsheet supergravity constraints (\ref{2dsugra})-(\ref{t-s}) are imposed. Then from (\ref{norma}) it follows that
\begin{equation}\label{chi0}
\chi^1_{\dot q}=\chi^2_{\dot q}=0
\end{equation}
and consequently 
\begin{equation}\label{3.39}
\psi^{1+}_{-2q}=\psi^{2-}_{+2q}=0,\ \psi^{1-}_{-2\dot q}=\psi^{2+}_{+2\dot q}=0.
\end{equation}
From Eqs.(\ref{pi1a+-cons}), (\ref{pi1a+2-2}) we derive that 
\begin{equation}\label{om=0}
\Omega^{+2i}_+=\Omega^{-2i}_-=0,
\end{equation}
\begin{equation}
\Omega^{+2i}_{+2}=\Omega^{-2i}_{-2}.
\end{equation}
When the worldsheet supergravity constraints are imposed Codazzi equations (\ref{PC}) allow to express $\Omega^{+2}_{+2}$ and $\Omega^{-2}_{-2}$ as the covariant derivatives of $\Omega^{+2}_{+}$, $\Omega^{-2}_{-}$
\begin{equation}
\Omega^{+2i}_{+2}=-\frac{i}{2}\nabla_+\Omega^{+2i}_+,\quad\Omega^{-2i}_{-2}=-\frac{i}{2}\nabla_-\Omega^{-2i}_-,
\end{equation}
so the nullification of $\Omega^{+2i}_+$ and $\Omega^{-2}_{-}$ (\ref{om=0}) signifies the nullification of 
\begin{equation}\label{om=0'}
\Omega^{+2i}_{+2}=\Omega^{-2i}_{-2}=0.
\end{equation}
In view of (\ref{3.39}), (\ref{om=0'}) we observe that the superstring equations of motion are satisfied, so the "minimal" embedding is on-shell. 

The only nonzero superfields within the "minimal" superembedding scheme are the chiral ones $\lambda^{1,2}_q(z^M)$ and $\psi^{1-}_{+2\dot q}(z^M)$, $\psi^{2+}_{-2\dot q}(z^M)$. So, the decompositions (\ref{p1+q}), (\ref{p2-q}), (\ref{p1-dotq}), (\ref{p2+dotq}) acquire the form
\begin{equation}
\pi^{1+}_q=-\frac{i}{2}e^{+2}\nabla_{+}\lambda^1_q+e^+\lambda^1_q,\ \pi^{2-}_q=-\frac{i}{2}e^{-2}\nabla_{-}\lambda^2_q+e^-\lambda^2_q,
\end{equation}
\begin{equation}
\pi^{1-}_{\dot q}=e^{+2}\psi^{1-}_{+2\dot q},\ \pi^{2+}_{\dot q}=e^{-2}\psi^{2+}_{-2\dot q}.
\end{equation}

In the end of the Section let us show how the relation between $\kappa-$symmetry gauge fixed physical variables of the NSR string and the Type IIA superstring arises. $e^+$, $e^-$ Components of the 1-form (\ref{apii}) read
\begin{equation}\label{nsr1}
\nabla_+x^i-\frac12\Omega^{+2i}_+x^{-2}-\frac12\Omega^{-2i}_+x^{+2}-i\chi^{1}_q\gamma^i_{q\dot q}\theta^{1-}_{\dot q}-ih^{1-}_{+\dot q}\tilde\gamma^i_{\dot qq}\theta^{1+}_{q}+ih^{2-}_{+q}\gamma^i_{q\dot q}\theta^{2+}_{\dot q}+i\chi^{2}_{\dot q}\tilde\gamma^i_{\dot qq}\theta^{2-}_{q}=0,
\end{equation}
\begin{equation}\label{nsr2}
\nabla_-x^i-\frac12\Omega^{+2i}_-x^{-2}-\frac12\Omega^{-2i}_-x^{+2}-ih^{1+}_{-q}\gamma^i_{q\dot q}\theta^{1-}_{\dot q}-i\chi^{1}_{\dot q}\tilde\gamma^i_{\dot qq}\theta^{1+}_{q}+i\chi^{2}_q\gamma^i_{q\dot q}\theta^{2+}_{\dot q}+ih^{2+}_{-\dot q}\tilde\gamma^i_{\dot qq}\theta^{2-}_{q}=0.
\end{equation}
The NSR string Grassmann variables $\psi^{\underline m}_{\pm}$ are the leading components of the worldsheet superfields $\Psi^{\underline m}_{\pm}=i\nabla_{\pm}X^{\underline m}$. In the superspace coordinate basis described in the Introduction we have for the transverse variables 
\begin{equation}\label{defphi}
\varphi^i_{\pm}=i\left(\nabla_{\pm}x^{i}-\frac12\Omega^{+2i}_{\pm}x^{-2}-\frac12\Omega^{-2i}_{\pm}x^{+2}\right).
\end{equation}
Substituting this expression back into (\ref{nsr1}), (\ref{nsr2}) and using (\ref{h0}), (\ref{chi0}) we derive
\begin{equation}
\varphi^i_+=\lambda^1_q\gamma^i_{q\dot q}\theta^{1-}_{\dot q},\ \varphi^i_-=-\lambda^2_q\gamma^i_{q\dot q}\theta^{2+}_{\dot q}, \end{equation}
which leading components as just the relations between the physical variables of the NSR string and the Type IIA superstring, as was argued in \cite{me}. 

\setcounter{equation}{0}
\section{Embedding of the $n=(8,8)$ superworldsheet into $D=~10$ $N=2B$ target superspace}

In the case under consideration basic equations governing the superembedding are
\begin{equation}\label{embb}
\Pi^{\pm2}=e^{\pm2},\ \Pi^i=0,
\end{equation}
\begin{equation}\label{embf}
\pi^{1+}_q=e^+_q,\ \pi^{2-}_{\dot q}=e^-_{\dot q}.
\end{equation}
The worldsheet superzweinbein definition $e^A=(e^{\pm2},e^+_q,e^-_{\dot q})$ (\ref{IIB})  differs from that adapted for the Type $IIA$ case (\ref{IIA}), its group structure is defined by irreducible $\kappa-$symmetry transformations, as was noted in the Introduction. 

The procedure of derivation of the integrability conditions for Eqs. (\ref{embb}), (\ref{embf}) is carried out among the lines of Section 3, so omitting intermediate formulae, we adduce the results. From (\ref{embb}) there follow the expressions for $\pm2$ components of the induced supertorsion 2-form
\begin{equation}\label{t+-28b}
T^{+2}=-2ie^{+}_{q}e^{+}_{q}-2i\pi^{2+}_{q}\pi^{2+}_{q},\quad T^{-2}=-2i\pi^{1-}_{\dot q}\pi^{1-}_{\dot q}-2ie^{-}_{\dot q}e^{-}_{\dot q}
\end{equation}
and the equation
\begin{equation}\label{dpii8b}
-\frac12\Omega^{+2i}e^{-2}-\frac12\Omega^{-2i}e^{+2}-2ie^{+}_q\gamma^i_{q\dot q}\pi^{1-}_{\dot q}-2i\pi^{2+}_q\gamma^i_{q\dot q}e^{-}_{\dot q}=0.
\end{equation}
From (\ref{embf}) one derives the expressions for $+q$, $-\dot q$ induced supertorsion components
\begin{equation}\label{tferm8b}
T^+_q=-\frac12\gamma^i_{q\dot p}\Omega^{+2i}\pi^{1-}_{\dot p},\quad T^-_{\dot q}=-\frac12\tilde\gamma^i_{\dot qp}\Omega^{-2i}\pi^{2+}_{p}.
\end{equation}
In (\ref{t+-28b}) we have used the definition (\ref{tors}) of the $\pm2$ components of the worldsheet supertorsion. Similarly to (\ref{torq}) we have defined $+q$, $-\dot q$ worldsheet supertorsion components as
\begin{equation}
T^{+}_{q}=de^{+}_{q}-\frac12\omega^{(0)}e^{+}_{q}+\frac14\omega_{qp}e^{+}_{p},\ T^{-}_{\dot q}=de^{-}_{\dot q}+\frac12\omega^{(0)}e^{-}_{\dot q}+\frac14\omega_{\dot q\dot p}e^{-}_{\dot p}.
\end{equation}
Besides that, like in Section 3, we have identified the induced and intrinsic connections
\begin{equation}
\Omega^{(0)}=\omega^{(0)},\ \Omega^{ij}=\omega^{ij}.
\end{equation}

Target superspace fermionic 1-forms $\pi^{1-}_{\dot q}$ and $\pi^{2+}_q$ that have not been identified with the superzweinbein can be expanded over the worldsheet basis as
\begin{equation}
\pi^{1-}_{\dot q}=e^{+2}\psi^{1-}_{+2\dot q}+e^{-2}\psi^{1-}_{-2\dot q}+e^+_ph^{1-}_{+p\dot q}+e^-_{\dot p}\chi^1_{\dot p\dot q},
\end{equation}
\begin{equation}
\pi^{2+}_{q}=e^{+2}\psi^{2+}_{+2\dot q}+e^{-2}\psi^{2+}_{-2\dot q}+e^+_p\chi^{2}_{pq}+e^-_{\dot p}h^{+2}_{-\dot pq}.
\end{equation}
The integrability conditions for these equations read
\begin{equation}\label{int8b}
\begin{array}{l} e^{+2}\nabla\psi^{1-}_{+2\dot q}+e^{-2}\nabla\psi^{1-}_{-2\dot q}+e^+_p\nabla h^{1-}_{+p\dot q}+e^-_{\dot p}\nabla\chi^1_{\dot p\dot q}+T^{+2}\psi^{1-}_{+2\dot q}+T^{-2}\psi^{1-}_{-2\dot q}+T^+_ph^{1-}_{+p\dot q}+T^{-}_{\dot p}\chi^1_{\dot p\dot q}\\[0.3cm]=-\frac12\tilde\gamma^i_{\dot qq}\Omega^{-2i}e^+_q,
\end{array}
\end{equation} 
\begin{equation}\label{int8b'}
\begin{array}{l}
e^{+2}\nabla\psi^{2+}_{+2q}+e^{-2}\nabla\psi^{2+}_{-2q}+e^+_p\nabla\chi^2_{pq}+e^{-}_{\dot p}\nabla h^{2+}_{-\dot pq}+T^{+2}\psi^{2+}_{+2q}+T^{-2}\psi^{2+}_{-2q}+T^+_p\chi^2_{pq}+T^-_{\dot p}h^{2+}_{-\dot pq}\\[0.3cm]=-\frac12\gamma^i_{q\dot q}\Omega^{+2i}e^-_{\dot q}.
\end{array}\end{equation}

The analysis of the derived integrability conditions we begin with the spinor-spinor components of Eq.(\ref{dpii8b}) since they are purely algebraic equations 
\begin{equation}\label{e+qe-pb}
\gamma^i_{q\dot q}h^{1-}_{+p\dot q}+\gamma^i_{p\dot q}h^{1-}_{+q\dot q}=0,\ 
\tilde\gamma^i_{\dot qp}h^{2+}_{-p\dot p}+\tilde\gamma^i_{\dot pp}h^{2+}_{-p\dot q}=0,\ 
\gamma^i_{q\dot q}\chi^1_{\dot q\dot p}+\tilde\gamma^i_{\dot pq}\chi^2_{pq}=0.
\end{equation}
To clarify the contents of (\ref{e+qe-pb}) let us decompose $h^{1-}_{+p\dot q}$, $h^{2+}_{-p\dot q}$, $\chi^1_{\dot q\dot p}$ and $\chi^2_{qp}$ over the full basis of the antisymmetric products of $SO(8)$ $\sigma-$matrices
\begin{equation}\label{b13} 
h^{1-}_{+p\dot q}=\gamma^i_{p\dot q}h^{1-i}_++\gamma^{ijk}_{p\dot q}h^{1-ijk}_+,\ h^{2+}_{-p\dot q}=\gamma^i_{p\dot q}h^{2+i}_++\gamma^{ijk}_{p\dot q}h^{2+ijk}_+,\ 
\end{equation} 
\begin{equation}\label{b14} 
\chi^1_{\dot q\dot p}=\chi^1\delta_{\dot q\dot p}+\chi^{1ij}\tilde\gamma^{ij}_{\dot q\dot p}+\chi^{1ijkl}_{\dot q\dot p},\ \chi^2_{qp}=\chi^2\delta_{qp}+\chi^{2ij}\gamma^{ij}_{qp}+\chi^{2ijkl}\gamma^{ijkl}_{qp}.
\end{equation} 
Plugging (\ref{b13}), (\ref{b14}) back into (\ref{e+qe-pb}) and performing straightforward manipulations with $\sigma-$matrices yields that
\begin{equation}\label{chi8b} 
h^{1-}_{+p\dot q}=h^{2+}_{-p\dot q}=0,\ \chi\equiv\chi^2=-\chi ^1,\ \chi^{1,2ij}=\chi^{1,2ijkl}=0.
\end{equation} 
The nonzero superfield $\chi(z^M)$ will be shown below to be the constant and related to the parameter of the global $SO(2)$ rotations of target superspace Grassmann coordinates $\Theta^{1,2\underline\mu}$. As is well known such rotations are the symmetry of the embedding equations rather than the Type IIB GS superstring action. Recently it was established that $\chi$ is related to the on-shell value of the field strength of $2d$ gauge field living  on the $D1-$brane, so that superembedding under consideration describes on equal footing both the Type IIB superstring and the $D1-$brane \cite{Bandos}.

The other components of Eq. (\ref{dpii8b}) read 
\begin{equation}\label{4.18}
\Omega^{-2i}_{-2}=\Omega^{+2i}_{+2},
\end{equation}
\begin{equation}
\Omega^{-2i}_{+q}=-4i\gamma^i_{q\dot p}\psi^{1-}_{+2\dot p},\ \Omega^{-2i}_{-\dot q}=-4i\tilde\gamma^i_{\dot qp}\psi^{2+}_{+2p},
\end{equation}
\begin{equation}\label{4.20}
\Omega^{+2i}_{+q}=-4i\gamma^i_{q\dot p}\psi^{1-}_{-2\dot p},\ \Omega^{+2i}_{-\dot q}=-4i\tilde\gamma^i_{\dot qp}\psi^{2+}_{-2p},
\end{equation}
provided (\ref{chi8b}) is used.

Then the nonzero $\pm2$ supertorsion components that emanate from (\ref{t+-28b}) are
\begin{equation}\label{tprbg}
T^{+2}_{+2-2}=4i\psi^{2+}_{+2q}\psi^{2+}_{-2q},\ T^{+2}_{\pm 2+q}=-4i\chi \psi^{2+}_{\pm2q},\ T^{+2}_{+q+p}= -4i\delta_{qp}(1+\chi^2);
\end{equation}
\begin{equation}
T^{-2}_{+2-2}=4i\psi^{1-}_{+2\dot q}\psi^{1-}_{-2\dot q},\ T^{-2}_{\pm 2-\dot q}=4i\chi \psi^{1-}_{\pm2\dot q},\ T^{-2}_{-\dot q-\dot p}= -4i\delta_{\dot q\dot p}(1+\chi^2).
\end{equation}
Similarly one infers from (\ref{tferm8b}) that 
\begin{equation}
T^{+q}_{+2-2}=\frac12\gamma^i_{q\dot p}\left(\Omega^{+2i}_{+2}\psi^{1-}_{-2\dot p}-\Omega^{+2i}_{-2}\psi^{1-}_{+2\dot p}\right),\ T^{+q}_{+p\pm2}=\frac12\gamma^i_{q\dot p}\Omega^{+2i}_{+p}\psi^{1-}_{\pm2\dot p},
\end{equation}
\begin{equation}
T^{+q}_{-\dot p\pm2}=\frac12\gamma^i_{q\dot q}\Omega^{+2i}_{-\dot p}\psi^{1-}_{\pm2\dot q}+\frac{\chi}{2}\gamma^i_{q\dot p}\Omega^{+2i}_{\pm2},
\end{equation}
\begin{equation}
T^{+q}_{+p-\dot p}=-\frac{\chi}{2}\gamma^i_{q\dot p}\Omega^{+2i}_{+p},\ T^{+q}_{-\dot q-\dot p}=-\frac{\chi}{2}\left(\gamma^i_{q\dot q}\Omega^{+2i}_{-\dot p}+\dot q\leftrightarrow\dot p\right),
\end{equation}
\begin{equation}
T^{-\dot q}_{+2-2}=\frac12\tilde\gamma^i_{\dot qp}\left(\Omega^{-2i}_{+2}\psi^{2+}_{-2p}-\Omega^{-2i}_{-2}\psi^{2+}_{+2p}\right),\ T^{-\dot q}_{-\dot p\pm2}=\frac12\tilde\gamma^i_{\dot qp}\Omega^{-2i}_{-\dot p}\psi^{2+}_{\pm2p},
\end{equation}
\begin{equation}
T^{-\dot q}_{+p\pm2}=-\frac{\chi}{2}\tilde\gamma^i_{\dot qp}\Omega^{-2i}_{\pm2}+\frac12\tilde\gamma^i_{\dot qq}\Omega^{-2i}_{+p}\psi^{2+}_{\pm2q},
\end{equation}
\begin{equation}\label{tpren}
T^{-\dot q}_{+p-\dot p}=\frac{\chi}{2}\tilde\gamma^i_{\dot qp}\Omega^{-2i}_{-\dot p},\ T^{-\dot q}_{+q+p}=\frac{\chi}{2}\left(\tilde\gamma^i_{\dot qq}\Omega^{-2i}_{+p}+q\leftrightarrow p\right).
\end{equation}
From the integrability conditions (\ref{int8b}), (\ref{int8b'}) it follows that 
\begin{equation}\label{8bzero}
d\chi=0,\ \psi^{1-}_{-2\dot q}=\psi^{2+}_{+2q}=0,\ \Omega^{+2i}_{+p}=\Omega^{-2i}_{-\dot p}=0,
\end{equation}
\begin{equation}
\psi^{1-}_{+2\dot q}=\frac{i}{32}\tilde\gamma^i_{\dot qp}\Omega^{-2i}_{+p},\ \psi^{2+}_{-2q}=\frac{i}{32}\gamma^i_{q\dot p}\Omega^{+2i}_{-\dot p},\ \nabla_{+p}\psi^{2+}_{-2q}=\nabla_{-\dot p}\psi^{1-}_{+2\dot q}=0,
\end{equation}
\begin{equation}\label{4.29}
\Omega^{+2i}_{+2}=\Omega^{-2i}_{-2}=\frac{4i\chi}{(1+\chi^2)}\psi^{1-}_{+2\dot q}\tilde\gamma^i_{\dot qp}\psi^{2+}_{-2p};
\end{equation}
\begin{equation}
\Omega^{-2i}_{+2}=-\frac{1}{4(1+\chi^2)}\nabla_{+q}\gamma^i_{q\dot p}\psi^{1-}_{+2\dot p},\ \Omega^{+2i}_{-2}=-\frac{1}{4(1+\chi^2)}\nabla_{-\dot p}\tilde\gamma^i_{\dot pq}\psi^{2+}_{-2q},
\end{equation}
where (\ref{chi8b}), (\ref{4.18})-(\ref{4.20}), (\ref{tprbg})-(\ref{tpren}) were utilized. Note, that in the presence of  parameter $\chi$ the superworldsheet mean curvature is nonzero. 

Because of (\ref{8bzero}) some components of the supertorsion (\ref{tprbg})-(\ref{tpren}) turn to zero, others reduce to
\begin{equation}
T^{+2}_{-2+q}=-4i\chi \psi^{2+}_{-2q},\ T^{+2}_{+q+p}= -4i\delta_{qp}(1+\chi^2);
\end{equation}
\begin{equation}
 T^{-2}_{+2-\dot q}=4i\chi \psi^{1-}_{+2\dot q},\ T^{-2}_{-q-p}= -4i\delta_{\dot q\dot p}(1+\chi^2);
\end{equation}
\begin{equation}
T^{+q}_{+2-2}=-\frac12\gamma^i_{q\dot p}\Omega^{+2i}_{-2}\psi^{1-}_{+2\dot p},
\end{equation}
\begin{equation}
T^{+q}_{-\dot p+2}=\frac12\gamma^i_{q\dot q}\Omega^{+2i}_{-\dot p}\psi^{1-}_{+2\dot q}+\frac{\chi}{2}\gamma^i_{q\dot p}\Omega^{+2i}_{+2},\ T^{+q}_{-\dot p-2}=\frac{\chi}{2}\gamma^i_{q\dot p}\Omega^{+2i}_{-2},
\end{equation}
\begin{equation}
 T^{+q}_{-\dot q-\dot p}=-\frac{\chi}{2}\left(\gamma^i_{q\dot q}\Omega^{+2i}_{-\dot p}+\dot q\leftrightarrow\dot p\right);
\end{equation}
\begin{equation}
T^{-\dot q}_{+2-2}=\frac12\tilde\gamma^i_{\dot qp}\Omega^{-2i}_{+2}\psi^{2+}_{-2p},
\end{equation}
\begin{equation}
T^{-\dot q}_{+p+2}=-\frac{\chi}{2}\tilde\gamma^i_{\dot qp}\Omega^{-2i}_{+2},\ T^{-\dot q}_{+p-2}=-\frac{\chi}{2}\tilde\gamma^i_{\dot qp}\Omega^{-2i}_{-2}+\frac12\tilde\gamma^i_{\dot qq}\Omega^{-2i}_{+p}\psi^{2+}_{-2q},
\end{equation}
\begin{equation}\label{tpren'}
T^{-\dot q}_{+q+p}=\frac{\chi}{2}\left(\tilde\gamma^i_{\dot qq}\Omega^{-2i}_{+p}+q\leftrightarrow p\right).
\end{equation}
We observe that like in the Type IIA case there survive only two chiral superfields $\psi^{1-}_{+2\dot q}(z^M)$ and $\psi^{2+}_{-2q}(z^M)$. The peculiarity of the Type IIB superembedding is the presence of the constant parameter $\chi$.

So, in the central basis for the target superspace coordinates one has
\begin{equation}\label{centr8b}
d\Theta^{1\underline\mu}=v^{-\underline\mu}_q\pi^{1+}_q+v^{+\underline\mu}_{\dot q}\pi^{1-}_{\dot q}=e^{+2}\psi^{1-}_{+2\dot q}v^{\underline\mu+}_{\dot q}+e^+_qv^{\underline\mu-}_q-\chi e^-_{\dot q}v^{\underline\mu+}_{\dot q},
\end{equation}
\begin{equation}\label{centr8b'}
d\Theta^{2\underline\mu}=v^{\underline\mu-}_{q}\pi^{2+}_{q}+v^{-}_{\underline\mu\dot q}\pi^{2+}_{\dot q},
=e^{-2}\psi^{2+}_{-2q}v^{\underline\mu-}_{q}+\chi e^+_qv^{\underline\mu-}_{ q}+e^{-}_{\dot q}v^{\underline\mu+}_{\dot q},
\end{equation}
\begin{equation}
\Pi^{\underline m}=u^{\underline m+2}e^{-2}+u^{\underline m-2}e^{+2}.
\end{equation}

Analogously to the Type IIA case we obtain the supersymmetric generalization of the Type IIB superstring equations of motion, that expressed in terms of the second fundamental form and $\psi-$superfields read 
\begin{equation}\label{gse2bf}
\psi^{1-}_{-2\dot q}(z^M)=\psi^{2+}_{+2q}(z^M)=0,
\end{equation}
\begin{equation}\label{gse2bb}
\Omega^{+2i}_{+2}(z^M)+\Omega^{-2i}_{-2}(z^M)+4i\left(\psi^{1+}_{+2q}(z^M)\gamma^i_{q\dot q}\psi^{1-}_{-2\dot q}(z^M)-\psi^{1+}_{-2q}(z^M)\gamma^i_{q\dot q}\psi^{1-}_{+2\dot q}(z^M)+1\leftrightarrow2\right)=0.
\end{equation}
From the fermionic embedding equations (\ref{embf}) it follows that $\psi^{1+}_{\pm2q}(z^M)=\psi^{2-}_{\pm2\dot q}(z^M)=0$, so (\ref{gse2bb}) transforms into 
\begin{equation}\label{gse2bb'}
\Omega^{+2i}_{+2}+\Omega^{-2i}_{-2}=0.
\end{equation}
It is easy to see that for the superstring (i.e., when $\chi=0$) equations (\ref{gse2bf}), (\ref{gse2bb'}) are satisfied as a result of (\ref{8bzero}), (\ref{4.29}). Thus, like in the case of Type IIA target superspace embedding of the $n=(8,8)$ superworldsheet is on-shell.

\setcounter{equation}{0}
\section{Embedding of the $n=(1,1)$ superworldsheet into $D=~10$ $N=2B$ target superspace}

In the previous Section when describing the superembedding of the $n=(8,8)$ worldsheet superspace we have identified $\pi^{1+}_q$ and $ \pi^{2-}_{\dot q}$ pieces of the target space basic fermionic 1-forms  (\ref{pialfa}) with the superworldsheet zweinbein fermionic components $e^+_q$, $e^-_{\dot q}$ (\ref{embf}). The motivation for this choice was the structure of the corresponding irreducible $\kappa-$symmetry transformations of the component formulation \cite{BZstring}. To describe the superembedding of the $n=(1,1)$ worldsheet superspace one needs to project $\pi^{1+}_q$ and $ \pi^{2-}_{\dot q}$ onto the bosonic $SO(8)-$spinor superfields $\lambda^1_q(z^M)$ and $ \lambda^2_{\dot q}(z^M)$ (\ref{embedfB2}) to adjust to fermionic components of the superzweinbein
\begin{equation}\label{embf1b}
\pi^{1+}_{q}\lambda^1_q=e^+,\ \pi^{2-}_{\dot q}\lambda^2_{\dot q}=e^-.
\end{equation}
Basic bosonic equations are just (\ref{embb}).

The derivation of the integrability conditions for equations describing the superembedding (\ref{embb}), (\ref{p1+q}), (\ref{p2-dotq}), (\ref{p2+q}), (\ref{p1-dotq'}), (\ref{embf1b}) proceeds in the same way as was described in the Section 3. From Eq.(\ref{embb}) one deduces the expressions for the induced supertorsion $\pm2$ components 
\begin{equation}\label{t+-21b}
T^{+2}=-2i\pi^{1+}_{q}\pi^{1+}_{q}\!-\!2i\pi^{2+}_{q}\pi^{2+}_{q},\quad T^{-2}=-2i\pi^{1-}_{\dot q}\pi^{1-}_{\dot q}\!-\!2i\pi^{2-}_{\dot q}\pi^{2-}_{\dot q}
\end{equation}
 and the equation
\begin{equation}\label{dpii1b}
-\frac12\Omega^{+2i}e^{-2}-\frac12\Omega^{-2i}e^{+2}-2i\pi^{1+}_q\gamma^i_{q\dot q}\pi^{1-}_{\dot q}-2i\pi^{2+}_q\gamma^i_{q\dot q}\pi^{2-}_{\dot q}=0
\end{equation}
that emanates from $\Pi^i=0$. Like everywhere through this paper we have indentified induced and intrinsic $SO(1,1)$ connections
\begin{equation}
\Omega^{(0)}=\omega^{(0)}
\end{equation}
when deriving (\ref{t+-21b}). 
The expressions for the $\pm$ components of the induced supertorsion originate as the integrability conditions for (\ref{embf1b})
\begin{equation}\label{t+B1}
T^+=\pi^{1+}_{q}\nabla\lambda^1_q-\frac12\lambda^1_q\gamma^i_{q\dot q}\Omega^{+2i}\pi^{1-}_{\dot q},
\end{equation}
\begin{equation}\label{t-B1}
T^-=\pi^{2-}_{\dot q}\nabla\lambda^2_{\dot q}-\frac12\lambda^2_{\dot q}\tilde\gamma^i_{\dot qq}\Omega^{-2i}\pi^{2+}_{q}.
\end{equation}
To derive Eqs.(\ref{t+-21b}), (\ref{t+B1}), (\ref{t-B1}) we have utilized the definition of $n=(1,1)$ worldsheet supertorsion (\ref{torpm2}), (\ref{torpm}). 

These equations should be supplemented by the integrability conditions for (\ref{p1+q}), (\ref{p2-dotq}), (\ref{p2+q}), (\ref{p1-dotq'}) that read
\begin{equation}\label{int1b'}\begin{array}{l}
e^{+2}\nabla\psi^{1,2+}_{+2q}\!+\! e^{-2}\nabla\psi^{1,2+}_{-2q}\!+\! e^+\nabla\chi^{1,2}_{q}\!+\! e^{-}\nabla h^{1,2+}_{-q}\!+\! T^{+2}\psi^{1,2+}_{+2q}\!+\! T^{-2}\psi^{1,2+}_{-2q}\!+\! T^+\chi^{1,2}_{q}\!+\! T^-h^{1,2+}_{-q}\\[0.3cm]=-\frac12\gamma^i_{q\dot q}\Omega^{+2i}\pi^{1(2)-}_{\dot q},
\end{array}\end{equation}
\begin{equation}\label{int1b}\begin{array}{l}
e^{+2}\nabla\psi^{1,2-}_{+2\dot q}\!+\! e^{-2}\nabla\psi^{1,2-}_{-2\dot q}\!+\! e^+\nabla h^{1,2-}_{+\dot q}\!+\! e^-\nabla\chi^{1,2}_{\dot q}\!+\! T^{+2}\psi^{1,2-}_{+2\dot q}\!+\! T^{-2}\psi^{1,2-}_{-2\dot q}\!+\! T^+h^{1,2-}_{+\dot q}\!+\! T^{-}\chi^{1,2}_{\dot q}\\[0.3cm]=-\frac12\tilde\gamma^i_{\dot qq}\Omega^{-2i}\pi^{1,2+}_q.
\end{array}\end{equation}
As a consistency check it is possible to verify that the multiplication of Eq.(\ref{int1b'}) for $\pi^{1+}_q$ by $\lambda^1_q$ and Eq.(\ref{int1b}) for $\pi^{2-}_{\dot q}$ by $\lambda^2_{\dot q}$ produces equations (\ref{t+B1}), (\ref{t-B1}). 

Let us analyse the consequences of the imposition of $2d$ supergravity constraints (\ref{2dsugra})-(\ref{t-s}) as we have done for Type IIA case (Section 3). The substitution of (\ref{2dsugra})-(\ref{t-s}) into the l.h.s. of (\ref{t+-21b}) and (\ref{t+B1}), (\ref{t-B1}) yields the following restrictions on the superfield coefficients of (\ref{p1+q}), (\ref{p2-dotq}), (\ref{p2+q}), (\ref{p1-dotq'})
\begin{equation}
h^{1,2+}_{-q}=h^{1,2-}_{+\dot q}=0,
\end{equation}
\begin{equation}\label{norma1b}
\chi^1_q\chi^1_q+\chi^2_q\chi^2_q=1,\  \chi^1_{\dot q}\chi^1_{\dot q}+\chi^2_{\dot q}\chi^2_{\dot q}=1.
\end{equation}
Equation (\ref{dpii1b}) defines the components of the second fundamental form via the coefficients of (\ref{p1+q}), (\ref{p2-dotq}), (\ref{p2+q}), (\ref{p1-dotq'})
\begin{equation}\label{pi1b+-cons}
\Omega^{-2i}_-=-4i\psi^{1+}_{+2q}\gamma^i_{q\dot q}\chi^1_{\dot q}-4i\psi^{2+}_{+2q}\gamma^i_{q\dot q}\chi^2_{\dot q},\quad \Omega^{+2i}_+=-4i\chi^1_q\gamma^i_{q\dot q}\psi^{1-}_{-2\dot q}-4i\chi^2_q\gamma^i_{q\dot q}\psi^{2-}_{-2\dot q},
\end{equation}
\begin{equation}\label{pi1bom}
\Omega^{+2i}_-=-4i\psi^{1+}_{-2q}\gamma^i_{q\dot q}\chi^1_{\dot q}-4i\psi^{2+}_{-2q}\gamma^i_{q\dot q}\chi^2_{\dot q},\quad \Omega^{-2i}_+=-4i\chi^{1}_{q}\gamma^i_{q\dot q}\psi^{1-}_{+2\dot q}-4i\chi^{2}_{q}\gamma^i_{q\dot q}\psi^{2-}_{+2\dot q}
\end{equation}
\begin{equation}\label{pi1b+2-2}
\Omega^{-2i}_{-2}-\Omega^{+2i}_{+2}=4i\psi^{1+}_{+2q}\gamma^i_{q\dot q}\psi^{1-}_{-2\dot q}-4i\psi^{1+}_{-2q}\gamma^i_{q\dot q}\psi^{1-}_{+2\dot q}+1\leftrightarrow2.
\end{equation}
and contains the algebraic equation
\begin{equation}\label{pi1b+-}
\chi^1_q\gamma^i_{q\dot q}\chi^1_{\dot q}+\chi^2_q\gamma^i_{q\dot q}\chi^2_{\dot q}=0,
\end{equation}
Finally there remains to contemplate  the integrability conditions (\ref{int1b'}), (\ref{int1b}). From (\ref{int1b'}) we find the spinor-spinor components 
\begin{equation}\label{psi1b}
\psi^{1,2+}_{+2q}=-\frac{i}{2}\nabla_+\chi^{1,2}_q,\quad \psi^{1,2+}_{-2q}=\frac{i}{4}\Omega^{+2i}_{-}\gamma^i_{q\dot p}\chi^{1,2}_{\dot p},
\end{equation}
\begin{equation}\label{5.18}
\nabla_-\chi^{1,2}_q=\frac12\Omega^{+2i}_+\gamma^i_{q\dot q}\chi^{1,2}_{\dot q}. 
\end{equation}
Eq. (\ref{int1b}) yields
\begin{equation}\label{psi1b'}
\psi^{1,2-}_{-2\dot q}=-\frac{i}{2}\nabla_-\chi^{1,2}_{\dot q},\quad \psi^{1,2-}_{+2\dot q}=\frac{i}{4}\Omega^{-2i}_+\tilde\gamma^i_{\dot qp}\chi^{1,2}_p,
\end{equation}
\begin{equation}\label{5.21}
\nabla_+\chi^{1,2}_{\dot q}=\frac12\Omega^{-2i}_-\tilde\gamma^i_{\dot qq}\chi^{1,2}_q.
\end{equation}
Spinor-vector and vector-vector component equations of (\ref{int1b'}), (\ref{int1b}) are the consequences of Eqs.(\ref{psi1b})-(\ref{5.21}).

Considered $n=(1,1)$ superworldsheet embedding is off-shell analogously to the Type IIA target superspace case. To see this we compare the Type IIB superstring fermionic equations of motion (\ref{gse2bf}) with Eqs. (\ref{psi1b}), (\ref{psi1b'})
\begin{equation}
\psi^{1-}_{-2\dot q}=-\frac{i}{2}\nabla_-\chi^{1}_{\dot q},\
\psi^{2+}_{+2q}=-\frac{i}{2}\nabla_+\chi^{2}_q
\end{equation}
and observe that the latter possess nonzero r.h.s.

As in Section 4 one is able to consider the "minimal" case when $\lambda^1_q$ and $\lambda^2_{\dot q}$ are identified with $\chi^1_q$ and  $\chi^2_{\dot q}$ respectively
\begin{equation}\label{iden1b}
\lambda^1_q=\chi^1_q,\  \lambda^2_{\dot q}=\chi^2_{\dot q},
\end{equation}
so that 
\begin{equation}
\lambda^1_q\lambda^1_q=\lambda^2_{\dot q}\lambda^2_{\dot q}=1.
\end{equation}
When additionally the worldsheet supergravity constraints are imposed it follows from (\ref{norma1b})  
\begin{equation}\label{chi01b}
\chi^2_q=\chi^1_{\dot q}=0.
\end{equation}
Because of (\ref{psi1b}), (\ref{psi1b'}) this leads to the nullification of $\psi^{1+}_{-2q}$, $\psi^{2+}_{+2q}$ and $\psi^{1-}_{-2\dot q}$, $\psi^{2-}_{+2\dot q}$
\begin{equation}
\psi^{1+}_{-2q}=\psi^{2+}_{+2q}=0,\ \psi^{1-}_{-2\dot q}=\psi^{2-}_{+2\dot q}=0.
\end{equation}
Equation (\ref{pi1b+-}) is then fullfilled identically. 

It is possible to show that "minimal" embedding is on-shell in the same way as was done in Section 4. 

We observe that there remain  four nonzero superfields: two bosonic $\lambda^1_q(z^M)$ and $\lambda^2_{\dot q}(z^M)$ and two fermionic $\psi^{2+}_{-2q}(z^M)$, $\psi^{1-}_{+2\dot q}(z^M)$. Thus target superspace fermionic 1-forms acquire the form 
\begin{equation}
\pi^{1+}_q=-\frac{i}{2}e^{+2}\nabla_{+}\lambda^1_q+e^+\lambda^1_q,\ \pi^{2-}_{\dot q}=-\frac{i}{2}e^{-2}\nabla_{-}\lambda^2_{\dot q}+e^-\lambda^2_{\dot q},
\end{equation}
\begin{equation}
\pi^{1-}_{\dot q}=e^{+2}\psi^{1-}_{+2\dot q},\ \pi^{2+}_{q}=e^{-2}\psi^{2+}_{-2\dot q}.
\end{equation}

Finally let us show the appearance of the relation between the $\kappa-$symmetry gauge fixed physical  variables of the NSR string and the Type IIB superstring. $+$ and $-$ components of equation $\Pi^i=0$ (\ref{embb}) read
\begin{equation}\label{nsr1b}
\nabla_+x^i-\frac12\Omega^{+2i}_+x^{-2}-\frac12\Omega^{-2i}_+x^{+2}-i\lambda^{1}_q\gamma^i_{q\dot q}\theta^{1-}_{\dot q}-ih^{1-}_{+\dot q}\tilde\gamma^i_{\dot qq}\theta^{1+}_{q}-(1\leftrightarrow 2)=0,
\end{equation}
\begin{equation}\label{nsr2b}
\nabla_-x^i-\frac12\Omega^{+2i}_-x^{-2}-\frac12\Omega^{-2i}_-x^{+2}-ih^{1+}_{-q}\gamma^i_{q\dot q}\theta^{1-}_{\dot q}-i\chi^{1}_{\dot q}\tilde\gamma^i_{\dot qq}\theta^{1+}_{q}-(1\leftrightarrow 2)=0.
\end{equation}
After substitution of (\ref{defphi}), (\ref{iden1b}), (\ref{chi01b}) expressions (\ref{nsr1b}), (\ref{nsr2b}) reduce to
\begin{equation}
\varphi^i_+=\lambda^1_q\gamma^i_{q\dot q}\theta^{1-}_{\dot q},\ \varphi^i_-=\lambda^2_{\dot q}\tilde\gamma^i_{\dot qq}\theta^{2+}_{q},
\end{equation}
which leading components are precisely the relations obtained in \cite{me} in the framework of the twistor-like Lorentz harmonic description of Type II superstrings.

\section{Conclusion}

We have considered possible mechanisms of generalization of the superembedding approach technique to the case when the number of the worldvolume Grassmann directions is less than $n_{ts}/2$, $n_{ts}$ being the number of the target space supersymmetries, on example of $n=(1,1)$ string superworldsheet embedded into $D=10$ Type II superspace. We formulated basic equations governing the superembedding and derived their integrability conditions. The consequences of the $2d$ $n=(1,1)$ supergravity constraints imposition on the worldsheet induced supertorsion were analysed. Unlike the on-shell $n=(8,8)$ superworldsheet embedding, the $n=(1,1)$ superworldsheet embedding is off-shell. There was also proposed the "minimal" superembedding scheme for which auxiliary spinor superfields are taken from the decomposition coefficients of the target space superforms over the superworldsheet. The imposition of the worldsheet supergravity constraints puts the "minimal' embedding on the mass shell. We have shown how the covariant relation between the physical variables of the NSR string and Type II superstrings originates in the framework of the "minimal" superembedding.

Construction of the corresponding worldsheet superfiled action that should generalize $D=3$ $N=2$ one of Ref. \cite{Galperin}, as well as, the extension of the proposed mechanisms of superembedding for diverse numbers of worldsheet supersymmetries and other branes can be the subject for future research. Another domain for exploration is a classification of such embeddings with $n_{wv}<n_{ts}/2$ according to a fraction of target space supersymmetries they preserve. 

\acknowledgments

The author is grateful to A.Yu. Nurmagambetov, D.P. Sorokin and A.A. Zheltukhin for the interest to the work and valuable discussions. The work was supported by Ukrainian State Foundation for Fundamental Research under Project 02.07/276.

\end{document}